\documentclass[journal]{IEEEtran}

\usepackage{glossaries}
\newacronym{ai}{AI}{All Intra}
\newacronym{hevc}{HEVC}{High Efficiency Video Coding}
\newacronym{avc}{AVC}{Advanced Video Coding}
\newacronym{jvet}{JVET}{Joint Video Experts Team}
\newacronym{cfp}{CfP}{Call for Proposal}
\newacronym{alf}{ALF}{Adaptive Loop Filter}
\newacronym{wf}{WF}{Wiener filter}
\newacronym{df}{DBF}{Deblocking Filter}
\newacronym{dsf}{DSF}{Diamond Shape Filter}
\newacronym{sao}{SAO}{Sample Adaptive Offset}
\newacronym{cc}{CCALF}{Cross-Component Adaptive Loop Filter}
\newacronym{aps}{APS}{Adaptation Parameter Set}

\newacronym{mpeg}{MPEG}{Motion Picture Experts Group}
\newacronym{vceg}{VCEG}{Video Coding Experts Group}

\newacronym{mse}{MSE}{Mean Squared Error}
\newacronym{tsmc}{TSMC}{Taiwan Semiconductor Manufacturing Company}

\newacronym{asic}{ASIC}{Application-Specific Integrated Circuit}
\newacronym{fpga}{FPGA}{Field Programmable Gate Array}
\newacronym{jem}{JEM}{Joint Exploration Model}
\newacronym{mts}{MTS}{Multiple Transform Selection}
\newacronym{lfnst}{LFNST}{Low Frequency Non-Separable Transform}
\newacronym{dct}{DCT}{Discrete Cosine Transform}
\newacronym{dst}{DST}{Discrete Sine Transform}
\newacronym{ctc}{CTC}{Common Test Conditions}
\newacronym{fft}{FFT}{Fast Fourier Transform}

\newacronym{simd}{SIMD}{Single Instruction Multiple Data}

\newacronym{sps}{SPS}{Sequence Parameter Set}
\newacronym{psp}{PSP}{Picture Parameter Set}
\newacronym{dsp}{DSP}{Digital Signal Processing}
\newacronym{dft}{DFT}{Discrete Fourier Transform}
\newacronym{qt}{QT}{Quad-Tree}
\newacronym{rdo}{RDO}{Rate Distortion Optimization}
\newacronym{vtm}{VTM}{VVC Test Model}
\newacronym{vhdl}{VHDL}{VHSIC Hardware Description Language}
\newacronym{alm}{ALM}{Adaptive Logic Module}
\newacronym{hw}{HW}{Hardware}
\newacronym{vb}{VB}{Virtual Boundary}
\newacronym{dpb}{DPB}{Decoded Picture Buffer}
\newacronym{mtt}{MTT}{Multi Type Tree}
\newacronym{rd}{RD}{Rate Distortion}
\newacronym{bt}{BT}{Binary Tree}
\newacronym{tt}{TT}{Ternary Tree}
\newacronym{cm}{CM}{Constant Multipliers}
\newacronym{tu}{TU}{Transform Unit}
\newacronym{fps}{fps}{Frame Per Second}
\newacronym{ssd}{SSD}{Sum of Squared Differences}
\newacronym{klt}{KLT}{Karhunen-Loéve Theorem}
\newacronym{nsst}{NSST}{Non-Separable Secondary Transform}
\newacronym{ra}{RA}{Random Access}
\newacronym{vvc}{VVC}{Versatile Video Coding}

\newacronym{iict}{IICT}{Inverse Integer Core Transforms}
\newacronym{vlsi}{VLSI}{Very Large Scale Integration}
\newacronym{bdr}{BD-BR}{Bj\o ntegaard Delta Rate}
\newacronym{sram}{SRAM}{Static Random Access Memory}

\newacronym{ram}{RAM}{Random-Access Memory}
\newacronym{rom}{ROM}{Read-Only Memory}
\newacronym{rm}{RM}{Regular Multiplier}
\newacronym{mcm}{MCM}{Multiple Constant Multiplier}

\newacronym{dc}{DC}{Design Compiler}
\newacronym{idst}{IDST}{Inverse DST}
\newacronym{idct}{IDCT}{Inverse DCT}

\newacronym{qp}{QP}{Quantization Parameter}
\newacronym{mc}{MC}{Motion Compensation}

\newacronym{cu}{CU}{Coding Unit}
\newacronym{ctu}{CTU}{Coding Tree Unit}
\newacronym{ctb}{CTB}{Coding Tree Block}
\newacronym{vvdec}{VVdeC}{Versatile Video deCoder}
\newacronym{esoc}{ESoC}{Embedded System-on-Chip}
\newacronym{lmcs}{LMCS}{Luma Mapping with Chroma Scaling}
\newacronym{lmp}{LMP}{Luma Mapping}
\newacronym{csp}{CSP}{Chroma Scaling}
\newacronym{cabac}{CABAC}{context adaptive binary arithmetic coding}
\newacronym{ilmcs}{ILMCS}{Inverse Luma Mapping Chroma Scaling}
\newacronym{sse}{SSE}{Streaming SIMD Extensions}
\newacronym{avx}{AVX}{Advanced Vector Extensions}
\newacronym{egpp}{EGPP}{Embedded general purpose processor}
\newacronym{wpp}{WPP}{Wave-front Parallel Processing}
\newacronym{simde}{SIMDe}{SIMDEverywhere}
\newacronym{ep}{EP}{Inter Prediction}
\newacronym{ip}{IP}{Intra Prediction}
\newacronym{tx}{TX}{Inverse Quantization and Transform}
\newacronym{ed}{ED}{Entropy Decoding}
\newacronym{cpu}{CPU}{Central Processing Unit}
\newacronym{gpu}{GPU}{Graphics Processing Unit}
\newacronym{hd}{HD}{High-definition}
\newacronym{fhd}{FHD}{Full High-definition}
\newacronym{uhd}{UHD}{Ultra High-definition}
\usepackage{amsmath,amsfonts}
\usepackage{algorithmic}
\usepackage{algorithm}
\usepackage{array}
\usepackage[caption=false,font=normalsize,labelfont=sf,textfont=sf]{subfig}
\usepackage{textcomp}
\usepackage{stfloats}
\usepackage{url}
\usepackage{verbatim}
\usepackage{graphicx}
\usepackage{cite}
\usepackage{adjustbox}
\usepackage{pgfplots}
\usepackage{textcomp}
\usepackage{multirow}
\usepackage{ragged2e}
\usepackage{footnote}
\usepackage{comment}
\usepackage{tikz}
\usepackage{pgf-pie}
\usepackage{caption}
\usepackage{soul}
\usepackage{booktabs} % To thicken table lines
\usepackage[para,online,flushleft]{threeparttable}

\ifCLASSINFOpdf
\else
\fi
\usepackage{hyperref}% Make Orcid icon
\definecolor{lime}{HTML}{A6CE39}
\DeclareRobustCommand{\orcidicon}{%
    \begin{tikzpicture}
    \draw[lime, fill=lime] (0,0) 
    circle [radius=0.16] 
    node[white] {{\fontfamily{qag}\selectfont \tiny ID}};    \draw[white, fill=white] (-0.0625,0.095) 
    circle [radius=0.007];    \end{tikzpicture}
    \hspace{-2mm}}
\foreach \x in {A, ..., Z}{%
    \expandafter\xdef\csname orcid\x\endcsname{\noexpand\href{https://orcid.org/\csname orcidauthor\x\endcsname}{\noexpand\orcidicon}}}

\begin{document}
\usetikzlibrary{patterns}

\title{Performance Analysis of Optimized Versatile Video Coding Software Decoders on Embedded Platforms}
    
%
%
%
% author names and IEEE memberships
% note positions of commas and nonbreaking spaces ( ~ ) LaTeX will not break
% a structure at a ~ so this keeps an author's name from being broken across
% two lines.
% use \thanks{} to gain access to the first footnote area
% a separate \thanks must be used for each paragraph as LaTeX2e's \thanks
% was not built to handle multiple paragraphs
%

\author{Anup Saha\orcidA{}, Wassim Hamidouche\orcidB{}, Miguel Chavarrías\orcidC{}, Guillaume Gautier\orcidD{}, Fernando Pescador\orcidE{},~\IEEEmembership{Senior Member,~IEEE,}, Ibrahim Farhat\orcidG{}
\thanks{Anup Saha, Miguel Chavarrías and Fernando Pescador were with CITSEM at Universidad Politécnica de Madrid, Madrid, Spain, e-mail: (anup.saha; miguel.chavarrias; fernando.pescador)@upm.es.

Wassim Hamidouche, Guillaume Gautier and Ibrahim Farhat were with Univ. Rennes, INSA Rennes, CNRS, IETR - UMR 6164, Rennes, France, e-mail: (wassim.hamidouche; guillaume.gautier; ibrahm.farhat)@insa-rennes.fr.}% <-this % stops a space
\thanks{This work was supported by both the Energy Efficient Enhanced Media Streaming (3EMS) project funded by the Brittany Region and TALENT project (PID2020-116417RB-C41), funded by the Spanish Ministerio de Ciencia y Innovación.}% <-this % stops a space
}

% note the % following the last \IEEEmembership and also \thanks - 
% these prevent an unwanted space from occurring between the last author name
% and the end of the author line. i.e., if you had this:
% 
% \author{....lastname \thanks{...} \thanks{...} }
%                     ^------------^------------^----Do not want these spaces!
%
% a space would be appended to the last name and could cause every name on that
% line to be shifted left slightly. This is one of those "LaTeX things". For
% instance, "\textbf{A} \textbf{B}" will typeset as "A B" not "AB". To get
% "AB" then you have to do: "\textbf{A}\textbf{B}"
% \thanks is no different in this regard, so shield the last } of each \thanks
% that ends a line with a % and do not let a space in before the next \thanks.
% Spaces after \IEEEmembership other than the last one are OK (and needed) as
% you are supposed to have spaces between the names. For what it is worth,
% this is a minor point as most people would not even notice if the said evil
% space somehow managed to creep in.

% The paper headers
\markboth{IEEE Transactions on Consumer Electronics, paper under review}%
{Shell \MakeLowercase{\textit{et al.}}: Bare Demo of IEEEtran.cls for IEEE Journals}
% The only time the second header will appear is for the odd numbered pages
% after the title page when using the twoside option.
% 
% *** Note that you probably will NOT want to include the author's ***
% *** name in the headers of peer review papers.                   ***
% You can use \ifCLASSOPTIONpeerreview for conditional compilation here if
% you desire.

% If you want to put a publisher's ID mark on the page you can do it like
% this:
%\IEEEpubid{0000--0000/00\$00.00~\copyright~2015 IEEE}
% Remember, if you use this you must call \IEEEpubidadjcol in the second
% column for its text to clear the IEEEpubid mark.

% use for special paper notices
%\IEEEspecialpapernotice{(Invited Paper)}

% make the title area
\maketitle

% As a general rule, do not put math, special symbols or citations
% in the abstract or keywords.

\begin{abstract}
In recent years, the global demand for high-resolution videos and the emergence of new multimedia applications have created the need for a new video coding standard. Hence, in July 2020 the \gls{vvc} standard was released providing up to 50\% bit-rate saving for the same video quality compared to its predecessor \gls{hevc}. However, this bit-rate saving comes at the cost of a high computational complexity, particularly for live applications and on resource-constraint embedded devices. This paper presents two optimized \gls{vvc} software decoders, named OpenVVC and \gls{vvdec}, designed for low resources platforms. They exploit optimization techniques such as data level parallelism using \gls{simd} instructions and functional level parallelism using frame, tile and slice-based parallelisms. Furthermore, a comparison in terms of decoding run time, energy and memory consumption between the two decoders is presented while targeting two different resource-constraint embedded devices. The results showed that both decoders achieve real-time decoding of \gls{fhd}  resolution over the first platform using 8 cores and \gls{hd} real-time decoding for the second platform using only 4 cores with comparable results in terms of average consumed energy: around 26 J and 15 J for the 8 cores and 4 cores platforms, respectively. Regarding the memory usage, OpenVVC showed better results with less average maximum memory consumed during run time compared to VVdeC.     
\end{abstract}

% Note that keywords are not normally used for peerreview papers.

\glsresetall

% For peer review papers, you can put extra information on the cover
% page as needed:
% \ifCLASSOPTIONpeerreview
% \begin{center} \bfseries EDICS Category: 3-BBND \end{center}
% \fi
%
% For peerreview papers, this IEEEtran command inserts a page break and
% creates the second title. It will be ignored for other modes.
\IEEEpeerreviewmaketitle

\section{Introduction} 
% The very first letter is a 2 line initial drop letter followed
% by the rest of the first word in caps.
% 
% form to use if the first word consists of a single letter:
% \IEEEPARstart{A}{demo} file is ....
% 
% form to use if you need the single drop letter followed by
% normal text (unknown if ever used by the IEEE):
% \IEEEPARstart{A}{}demo file is ....
% 
% Some journals put the first two words in caps:
% \IEEEPARstart{T}{his demo} file is ....
% 
% Here we have the typical use of a "T" for an initial drop letter
% and "HIS" in caps to complete the first word.
\IEEEPARstart{A}{}new era of information and communication technologies is emerging, where video communication plays an essential role in internet traffic. In particular, the significant increase in video traffic supported by the COVID19 global health situation and the emerging video formats and applications have lead to the development of a new video coding standard, named \gls{vvc}/H.266. This latter was standardized in July 2020 by the \gls{jvet} of the VCEG working group of ITU-T and the MPEG working group of ISO/IEC JTC 1/SC 29~\cite{vvcwiki}. \gls{vvc} enables bit-rate savings of up to 50\% \cite{fvvc} with respect to the previous standard \gls{hevc}/H.265 \cite{hevc2013} for the same video quality. However, this achievement comes at the cost of $10\times$ and $2\times$ more complexity compared to \gls{hevc} for the encoder and decoder contexts, respectively \cite{VVCComplexity}. In this scenario, the main challenge is to develop real-time \gls{vvc} codecs that take into account resource-constrained consumer devices frequently used in consumer electronics based on embedded platforms. 

Each coding standard is released with a reference software implementation available for the scientific community. These solutions incorporate all the basic features of the standard but offer a very limited speed performance. In the case of \gls{vvc}, the reference software is the \gls{vtm} \cite{vtm_ref}. Taking this as a starting point, research groups and companies develop their own real time software and hardware solutions. Depending on the targeted architecture, these solutions mainly exploit the intrinsic parallelism of the algorithms, both at the data and functional levels, to enhance their performance in terms of speed and energy consumption. In the first case, some data operations included in the source code are optimized by using \gls{simd} type instructions \cite{SIMD_other3}. Here, vectorized operations are used to perform mathematical operations with more than one operator using a single processor instruction. The other potential optimisation route is to take advantage of the intrinsic parallelism of independent processing of pictures \cite{frame}, or smaller parts of the picture, such as slices \cite{slice} or tiles \cite{tile}. In the latter case, it is necessary that the coding is done by activating these normative tools that break dependencies between adjacent regions.    

In this work, two open source \gls{vvc} decoders are presented and compared against each other. These solutions, named OpenVVC \cite{openvvc1, openvvc} and \gls{vvdec} \cite{vvdec} decoders, are optimized using data and functional level parallelism techniques. In this paper, we evaluate their performance in terms of decoding run time, power consumption and memory usage targeting two different embedded platforms. The results showed that both decoders have achieved 15 to 34 \gls{fps} for \gls{uhd} sequences with \gls{qp} 27 and 37, and achieved real-time decoding of \gls{fhd} and \gls{hd} sequences over the first target platform using 8 cores. Furthermore, 16 to 28 \gls{fps} have been obtained for \gls{fhd} sequences with \gls{qp}s 27 and 37, and real-time  decoding has been reached for all \gls{hd} sequences by OpenVVC and \gls{vvdec} when targeting the second embedded platform with 4 cores. In terms of energy consumption and maximum memory usage, the experimental results showed that \gls{vvdec} has consumed $\times 2$ more memory with respect to the consumption of OpenVVC over both target platforms. On the other hand, OpenVVC consumed the same energy as \gls{vvdec} on the \gls{esoc}1 platform with 8 cores and around 1.25$\times$ \gls{vvdec}'s energy consumption when targeting \gls{esoc}2 embedded platform with 4 cores.  

The rest of the paper is structured as follows. Section \ref{background} give a short introduction to the \gls{vvc} standard. Section \ref{RTDecoders} describes the optimizations included in the \gls{vvc} decoders using specific parallelization techniques along with the state-of-the-art of \gls{vvc} decoders. Section \ref{decoder optimisation} details the proposed optimizations techniques included in the OpenVVC decoder. The obtained results and comparison between OpenVVC and \gls{vvdec} performance are provided in Section \ref{result}. Finally, Section \ref{conclusion} concludes the paper.

%\hl{Update this section} The rest of the paper is structured as follows. Section \ref{background} presents the background of the \gls{vvc} standard along with the state-of-the-art of \gls{vvc} decoders. Section \ref{RTDecoders} describes the optimizations included in the VVC decoders in terms of parallelization of the algorithm to speedup the decoding process. Section \ref{decoder optimisation} details the proposed optimizations techniques included in the OpenVVC decoder. The obtained results and comparison between OpenVVC and \gls{vvdec} performance are provided in Section \ref{result}. Finally, Section \ref{conclusion} concludes the paper.

%\section{Background} \label{background}
%This section presents the most relevant related work available in the literature. In particular, two open source software implementations compliant with the VVC standard, OpenVVC and VVdeC, are described briefly. Lastly, a description of the parallel processing techniques used to speedup the performance of software implementation of video decoders is provided \hl{it is not what described}. 

\section{Introduction to VVC} \label{background}
\begin{figure}[b]
\centering
\includegraphics[width=0.5\textwidth, height= 0.35\textwidth]{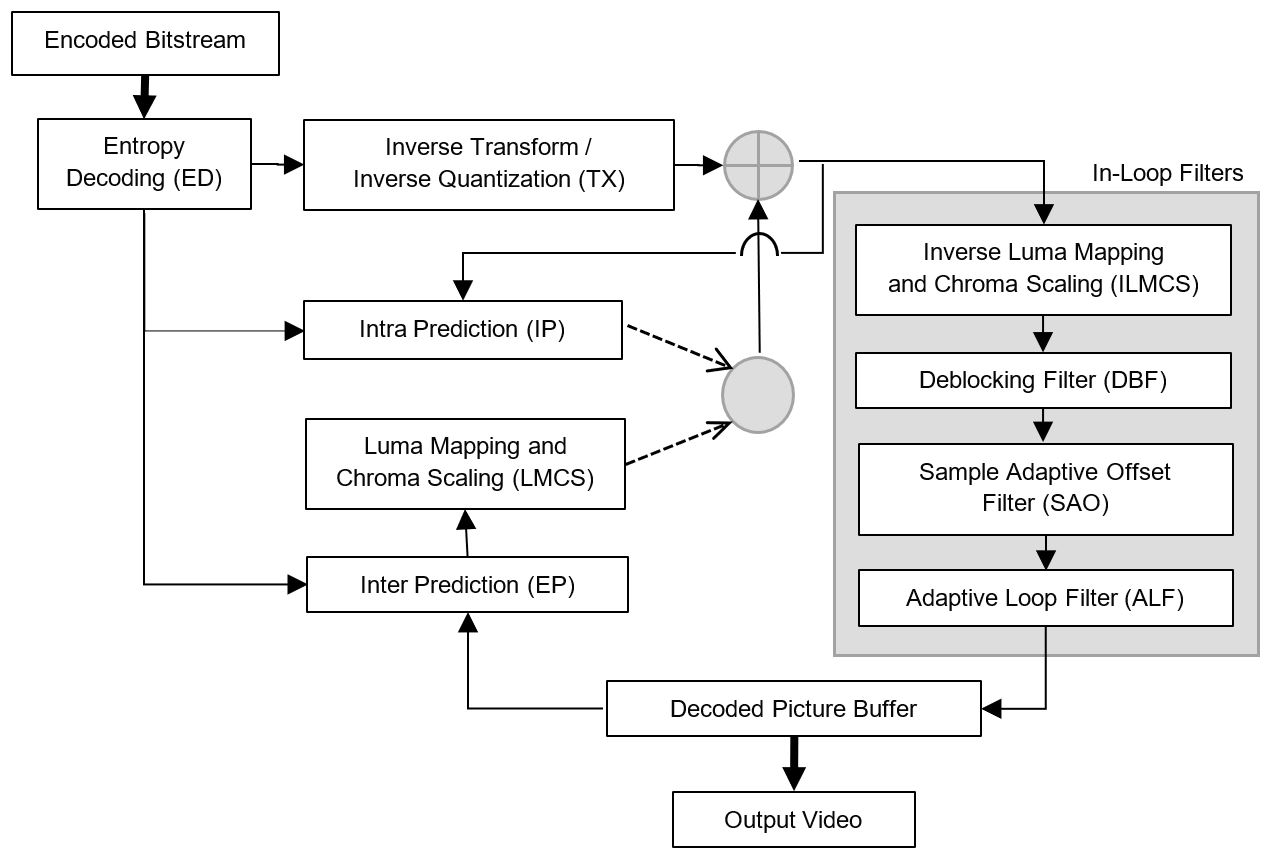}
\caption{Block diagram of a \gls{vvc} decoder.}
\label{vvcdeco}
\end{figure}

Similar to its predecessors, \gls{vvc} was designed based upon a hybrid coding scheme using intra/inter-prediction coding and transform coding. In Figure \ref{vvcdeco}, the decoding process scheme of \gls{vvc} is presented. Here, the encoded bit-stream is the input, the decoded video is the output, and each decoding phase is presented by one block.

\subsection{Entropy decoding} Bit-stream decoding begins with this process integrated with similar but advanced \gls{cabac} \cite{cabac} with respect to \gls{hevc}. Here, an updated multihypothesis probability estimation model was adapted and the computed look-up table was eliminated for enhancing the accuracy. The coefficient coding has been improved by allowing 1$\times$16, 2$\times$8, 8$\times$2, 2$\times$4, 4$\times$2, and 16$\times$1 coefficients group size for transform block size. 

\subsection{Inverse quantization and transform}
The spatial domain coefficients are retrieved from the frequency domain by inverse quantization and inverse transformation. \gls{vvc} introduces \gls{mts} \cite{mts7} tool used to encode the residual inter and intra coding blocks. \gls{mts} allows three transforms of the rectangle blocks with the height and width of $\leq64$ for \gls{dct}-II, $\leq32$ for \gls{dct}-VIII and \gls{dst}-VII. Moreover, the coefficients of high frequency are zeroed when the height and or width is equal to 64 for \gls{dct}-II and 32 for \gls{dct}-VIII and \gls{dst}-VII. In addition, \gls{lfnst} \cite{ilfnst} is used on the low frequency coefficients of the transform domain after \gls{dct}-II for further signal decorrelation.

\subsection{Intra prediction}
In \gls{vvc} intra-prediction module, 32 additional directional intra-prediction modes were added with respect to \gls{hevc}. Moreover, it allows a wide-angle intra modes for rectangular blocks, which improves the prediction accuracy. In addition, the matrix weighted intra-prediction tool was used a new intra prediction mode by taking above and left neighbouring lines of the prediction block. Besides, \gls{vvc} adapted cross-component linear model \cite{cclm} tool, which applied to the prediction of chroma components from the luma components.   

\subsection{Inter prediction}
Inter-prediction takes advantage of the temporal redundancy of the video by removing the correlation in the temporal domain \cite{alfg}. Here, the motion compensation estimates the current coding unit samples according to the samples recorded in the decoded picture buffer. In addition, 8-tap filter is used to luma samples for creating motion-compensated prediction and 4-tap filter is used to chroma samples for interpolation \cite{vvcom}. Furthermore, \gls{vvc} achieved improved prediction accuracy using decoder-side motion vector refinement\cite{dmvr} and bi-directional optical flow prediction refinement \cite{bdof}.  

\subsection{Luma mapping with chroma scaling}
Forward \gls{lmcs} is a new tool introduced in \gls{vvc} and comes after inter-prediction stage. It has two parts: \gls{lmp}, used to modify luma predicted samples, and \gls{csp}, used to modify chroma residues. \gls{lmp} makes the most use of the range of luma code values and provides an efficient reallocating process of luma code values in the coding domain. Therefore, \gls{csp} changes the value of the chroma residual samples in the chroma coding block for mitigating the defect coming from the interaction between luma and luma corresponding chroma signals \cite{lmcs}. %The LMCS detailed processing is depicted in \cite{lmcs}, \cite{vil}.

\subsection{In-loop Filters}
\gls{vvc} in-loop filters consist of inverse \gls{ilmcs} \cite{vil}, \gls{df}, \gls{sao} filter and \gls{alf}. First, \gls{ilmcs} is a new addition to \gls{vvc} which enhances the decoding performance by inverse mapping the luma code to the reconstructed block. \gls{df} and \gls{sao} in \gls{vvc} are very similar to \gls{hevc} \cite{comv}. \gls{df} is used to detect and filter the artifacts of the pixels at the block boundaries and \gls{sao} is used to minimize sample distortion over the pixels filtered by \gls{df}. In addition, unlike \gls{hevc}, \gls{vvc} includes \gls{alf} \cite{vil} for reducing mean-squared error of the decoded pixels. Therefore, undesired artifacts obtained by the previous decoding modules including blurring, blocking, flickering, ringing, and colour shift are mitigated using in-loop filters and the decoded video is obtained.

\section{Optimized and real time software decoders} \label{RTDecoders}
There are three levels of parallelism that can be exploited to speedup the video decoding process: data-level, frame-level and high-level. In this section we first describe these three levels of parallelism, then we give a brief description of existing optimised decoders with an introduction to the recent software and real time \gls{vvc} decoders including OpenVVC and \gls{vvdec}.   
% In turn, frame-level exploits multi-threads to process multiple frames in parallel by fulfilling dependencies of the motion compensation. Second, high-level parallelism comprises two possible types: normative and non-normative. The most used normative high-level parallelisms are wave-front parallel processing and tiles parallelism. Finally, data-level parallelism takes advantage of the possibility to vectorice some mathematical operations using SIMD instructions, among others. Frame-level, WPP, tiles, and SIMD parallelism is described in this section. 

\subsection{Task level parallelism}

\subsubsection{Single instruction multiple data} 
\gls{simd} is a data-level parallel processing technique that loads multiple data in single register to operate with. The x86-based architectures offer \gls{sse} and \gls{avx} based \gls{simd} intrinsics. \gls{egpp} based platforms support ARM Neon suite instead of \gls{sse} and \gls{avx} based \gls{simd} intrinsic. ARM Neon is an advanced \gls{simd} technology designed for mobile devices that support up to 128-bit register. Neon-based \gls{simd} technology can be used following 4 ways \cite{armneon}: a) using Neon intrinsics, b) Neon-enabled libraries, c) compiler auto-vectorization, and d) hand-coded Neon assembler.

\subsubsection{Frame-level parallelism}
Simultaneous processing of multi-frames is performed in frame-level parallelism while dependencies of the motion compensation are also satisfied. The length of the motion vector is, in this case, the deterministic factor \cite{slice}. Video sequences with large motion would imply large dependencies between frames which may create a major disadvantage for frame-level parallelism. Hence, sequences in all intra configuration are the most benefited by frame-level parallelism due to the fact that there are no motion compensation dependencies. Moreover, in frame-level parallelism, additional picture buffer and local buffers should be stored for each thread used to decode in parallel. As a result, it demands higher memory than sequential decoding.

\subsection{High level parallelism}

\subsubsection{Wave-front parallel processing}
\gls{wpp} allows virtual picture partitioning into \gls{ctu} rows \cite{slice}. In \gls{wpp}, the coding dependency is kept unchanged during the picture partitioning while the entropy engine is initialized at the start of each \gls{ctu} line. Therefore, at the beginning of each \gls{ctu} row, the \gls{cabac} context is re-intialized and it depends at least on the data from first \gls{ctu} of the previous row. As a result, the decoding of rows could not completed at the same time and slightly limits parallelisation efficiency of \gls{wpp}.   

\subsubsection{Tiles parallelism}
\begin{figure}[hbt!]
\centering
\includegraphics[width=0.3\textwidth, height= 0.2\textwidth]{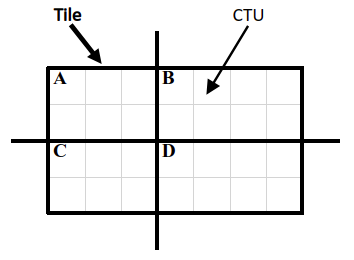}
\caption{Illustration of tile partitioning in VVC decoder.}
\label{tilef}
\end{figure}

\gls{vvc} supports tiles of rectangular shape consisting of \gls{ctu}s \cite{til}. An example of tile partition is shown in Fig. \ref{tilef}. Here, four tiles are labelled with A, B, C, and D. Tiles are separated by boundaries, which eliminates the prediction dependencies. Therefore, for all the tiles, the entropy encoding step is reinitialized, which allows to decode tiles independently. It allows decoding a picture concurrently using multiple threads. However, the in-loop filtering process can only be carried out at the tile boundaries when pixels are reconstructed from both sides. 

\subsection{Codec optimisations} 
Various parallelization methods have been used to accelerate video codecs on \gls{cpu}, \gls{gpu}, and combination of both based architectures. Yan et al. accelerated a \gls{hevc} decoder by $\times4$ compared to HM 4.0 using \gls{simd} technologies over a x86 processor in \cite{lyjz}. Authors in \cite{uls1} and \cite{uls2} proposed \gls{gpu}-based implementation of \gls{hevc} that satisfied real-time requirements for the decoding of \gls{uhd} 4k sequences. S. Gudumas et al. in \cite{slice} discussed various optimization techniques to implement \gls{vvc} using multiple \gls{cpu} cores on heterogeneous platforms with the purpose of achieving real-time decoding \cite{slice}. Here, the decoding tasks were redesigned and parallelized with task parallelization based on load balancing and data parallelization, at the level of \gls{ctu}. The authors in \cite{gpuxssw} presented a \gls{gpu}-based motion compensation system to accelerate the \gls{vvc} decoder that exploits the partition of the different \gls{cu} and proper thread organisation. Furthermore, Wieckowski et al. in \cite{fvvc} described an optimized \gls{vvc} decoder that achieved real-time decoding on general purpose \gls{cpu}s. Here, \gls{simd} operations-based optimisation and multi-threading based optimisation were adapted. The authors in \cite{binZhu} demonstrated an optimized real-time \gls{vvc} decoder that takes advantage of \gls{simd} instruction extensions on x86 based \gls{cpu}. Moreover, the authors discussed the implementation of frame, \gls{ctu}, sub-\gls{ctu}, and task level parallelization. An optimized \gls{vvc} software decoder on mobile platforms is presented in \cite{alfg},. The presented decoder was generated from \gls{vtm}-11.0 reference software. Here, the decoder achieved real-time decoding for HD video sequences using \gls{simd} and multi-threading on ARM \cite{comv} based platform.

%\subsection{Optimized VVC decoders}
%\hl{May ve section C and D can be merged in one section} 
There are a handful of software open-source decoders available that are compliant with the \gls{vvc} standard. Firstly, the already mentioned reference \gls{vvc} test model, or \gls{vtm}. Secondly, \gls{vvdec} \cite{vvdec}, an implementation proposed by Fraunhofer Institute, is an optimized decoder based on \gls{vtm}. It includes \gls{simd} and multi-threads parallelisation for an optimal performance. Thirdly, O266dec \cite{o266}, a real time decoder proposed by Tencent Media Lab. O266dec also uses the \gls{simd} and multithreading parallelisation techniques. Unfortunately this decoder has not been updated in the last 14 months. Finally, OpenVVC is a lightweight open source software decoder that is available in \cite{openvvc}. It is designed to target different operating systems and hardware architectures. Similar to \gls{vvdec} and O266dec, OpenVVC uses data and functional level parallelism to optimize the decoding performance. For more details on \gls{vvc} codecs, Sullivan
~\cite{VVCstatus} provides a complete list of available \gls{vvc} encoder and decoder implementations. 
%Other state-of-the-art VVC compliant decoder proposed by Tencent Media Lab is O266dec \cite{o266}. O266dec also uses the SIMD and multithreading parallelisation techniques. Finally, OpenVVC is a light VVC decoder that is openly available in \cite{openvvc}. It is designed for different operating systems and architectures. In the following section, OpenVVC and VVdeC are described.    

\subsubsection{Introduction to OpenVVC} 
OpenVVC is an open source software \gls{vvc} decoder written in C programming language. It is compiled as a cross-platform library and is compatible with most-used operating systems and optimized for x86 and ARM processors. The last version of the decoder is compliant with \gls{vvc} Main profile. In addition, it is integrated with VLC \cite{vlc}, GPAC \cite{gpac}, and FFplay \cite{ffplay} video players. OpenVVC provides high decoding speed by using as little memory as possible. It exploits tiles and frames paralelization based on multi-CPU cores, and \gls{simd} optimisations for accelerating the decoding process. The decoding process of OpenVVC starts by parsing the parameters of the sequence. Therefore, reconstruction tasks consisting of \gls{tx}, \gls{lmcs}, \gls{ep} and \gls{ip} decoding blocks, are performed at the \gls{cu} level. Then, \gls{df} is performed immediately after the reconstruction process is completed at the level of \gls{ctu}. This approach helps to optimize memory usage by avoiding massive storage of quantization parameter map and \gls{cu} dimension that is essential for the \gls{df} process. Finally, \gls{alf} is applied after the \gls{sao} at the level of \gls{ctu} line and this process delivers the decoded frame as output. 

\subsubsection{Introduction to Versatile Video Decoder}
\gls{vvdec} \cite{vvdec} is an open source software \gls{vvc} decoder optimized for x86 architectures and developed by Fraunhofer Institute for Telecommunications, Heinrich Hertz Institute, HHI. Unlike OpenVVC, \gls{vvdec} has been developed from \gls{vtm} reference software~\cite{2vvdec}. It supports \gls{vvc} Main 10 profile and it is capable to decode all conformance \gls{vvc} bitstreams \cite{3vvdec}. In addition, \gls{vvdec} comes with \gls{simd} optimisations and multi-threading parallelisation over x86 architectures. The parallelisation of \gls{vvdec} decoding begins by parsing multiple frames concurrently. Therefore, in the reconstruction process, tasks are split based on \gls{ctu} lines and \gls{ctu}s. Here, a stage is given to each \gls{ctu} for tracking the following stage and process tasks in parallel after the dependencies are satisfied. It allows task coordination where a task worker is assigned to each \gls{ctu}. Task workers are scanned by a thread-pool for assigning the available task. \gls{vvdec} has achieved decoding time reduction up to 90\% \cite{fvvc} with respect to \gls{vtm}.      

\section{Decoder Optimizations}\label{decoder optimisation}
This section describes the implementation of frames, tiles, and Neon-based \gls{simd} parallelisms in OpenVVC over \gls{egpp}-based platforms. 

\subsection{Frames and Tiles parallelization in OpenVVC}
In frame-level parallelism of OpenVVC, a main thread is used to parse the \gls{psp}, \gls{sps}, picture/slice header and schedule decoding threads with a thread pool. Then, the main thread provides the data and updates the internal structure of the available threads in the thread-pool for decoding the frame. Therefore, motion compensation synchronisation between threads is performed for sequences with inter-coding configuration after starting the decoding process. In fact, this latter is the most challenging step in frame-level parallelism, where the available thread has to wait for motion compensation before starting the pixels processing. When pixels are ready, the available thread is able to perform the decoding process. This process is applied at the \gls{ctu} line level since OpenVVC performs decoding and in-loop filtering at \gls{ctu} line basis. Once the decoding process is completed, the decoding threads signal their availability to the main thread and return to the thread-pool. 

On the other hand, tiles level parallelism is applied at a portion of a frame. In fact, every frame is decomposed into rectangular regions of the picture containing multiple \glspl{ctu}~\cite{til}. The main challenge of tile level parallelism is that tiles could have different run time complexities. Therefore, the time required to finish one frame is the time to finish the longest tile. In this case, at a certain processing time, some threads are free, without a task, waiting to finish processing the current frame. For more details about this issue, a quality-driven dynamic frame partitioning for efficient parallel processing is explained by Amestoy et al. in \cite{Amestoy}. A dynamic tile and rectangular slice partitioning solution enables the best partitioning configuration that minimizes the trade-off between multi-thread encoding time and quality loss. This is performed by taking into account both spatial texture of the content and encoding times of previously encoded frames. Experiments prove that the proposed solution, compared to uniform partitioning, significantly decreases multi-thread encoding time, with slightly better quality.

\begin{figure}[hbt!]
\centering
\includegraphics[width=0.4\textwidth, height= 0.4\textwidth]{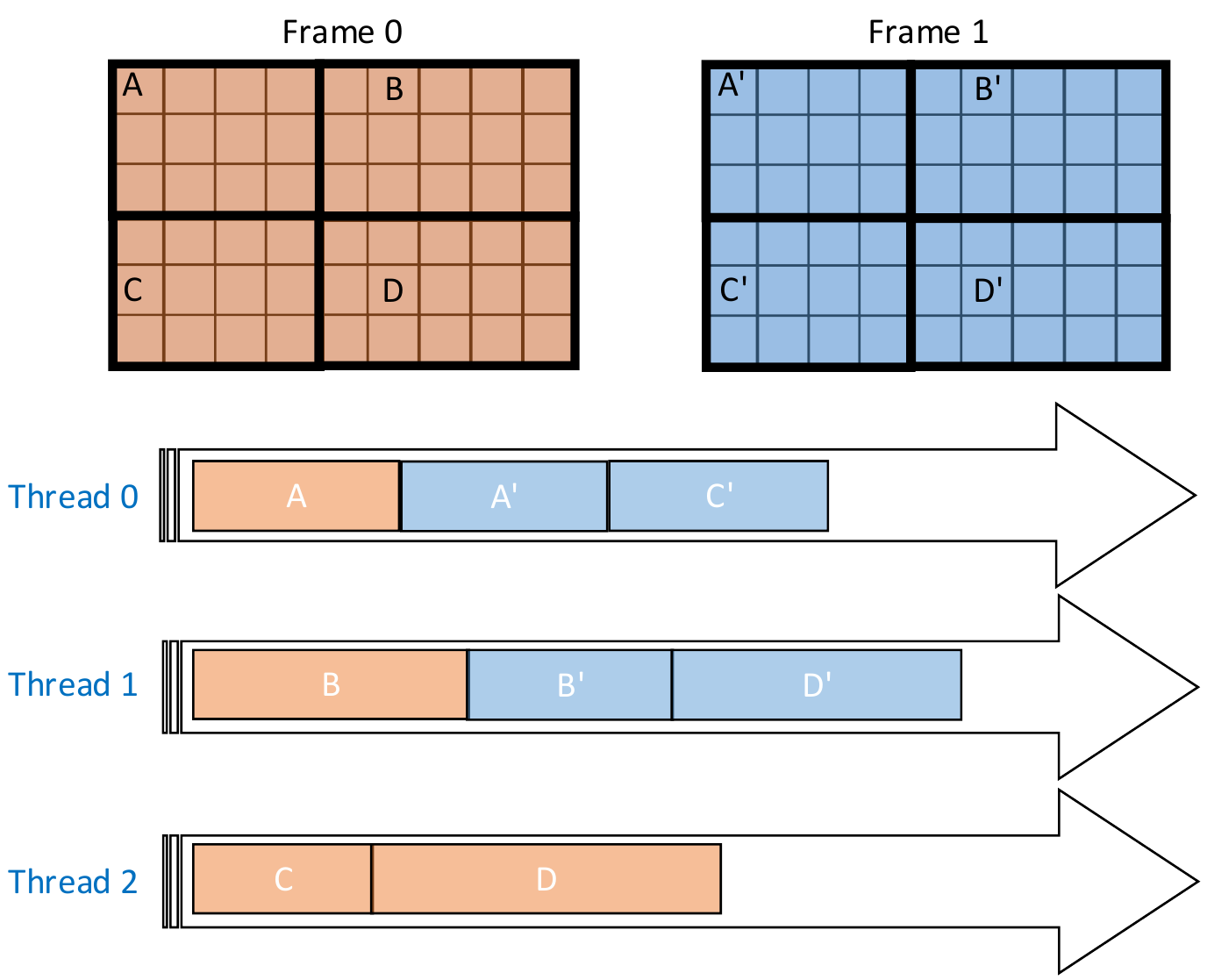}
\caption{Illustration of tile pipelining in OpenVVC decoder.}
\label{tilePipeline}
\end{figure}

The proposed solution integrated in OpenVVC aims to efficiently activating all threads at all times. In order to do so, it applies thread pipelining technique that overlooks frames and focuses only on tiles. Figure~\ref{tilePipeline} illustrates tile pipelining. The tile partitioning forms a 2$\times$2 grid. They are labeled  A, B, C and D for the first frame and A', B', C' and D' for the second frame and delimited by the thicker black lines. Prediction dependencies across tile boundaries are broken and entropy encoding state is reinitialized for each tile. These restrictions ensure that tiles are independently decoded, allowing several threads to decode simultaneously the same picture. As it can be observed, regardless of the tile position, as soon as a thread is available from the thread pool, the tile is processed. Thread 2 for example does not work on any tiles of the second frame since it took the entire time working on tile D of the first frame. This fact does not restraint threads 0 and 1 from working on the tiles of the second frame. However, adopting this technique creates some sort of a combination between frame and tile parallelism, as a result, dependencies between frames for inter prediction and motion compensation should be taken into account. This latter is handled by OpenVVC. Moreover, since OpenVVC processes tiles independently of their frame affiliation, tile size and load optimization at the encoder side do not actually impact the performance of OpenVVC. At the end, tiles are pipelined regardless of their size or load without waiting to finish processing the current frame.

\subsection{SIMD optimisation in OpenVVC} 
In this study, ARM Neon-based \gls{simd} optimisations were adapted to accelerate OpenVVC targeting \gls{egpp}-based platforms. First, the x86 architecture-based \gls{simd} intrinsics used in OpenVCC are replaced by the ARM Neon-based \gls{simd} intrinsics. Therefore, additional adjustment was adapted due to the fact that Neon-based intrinsics are not as powerful and complete as compared to \gls{sse} or \gls{avx} intrinsics. In particular, for some cases, one \gls{simd} instruction for x86-based was replaced with multiple Neon-based \gls{simd} instructions. For instance, two Neon intrinsics vmull\_s16 for multiply operation and vpaddq\_s32 for add operation are needed to replace madd\_epi16.  

\gls{esoc}s used in this study support up to 128-bit \gls{simd} registers. A 128-bit register can be loaded with 16 8-bit, 8 16-bit, 4 32-bit, or 2 64-bit data. This fact allows concurrent data processing to achieve a theoretical speedup of up to $\times16$ on 8-bit data. In this study, Neon-based \gls{simd} instructions are used to optimize the high computational demanding \gls{vvc} decoder modules presented in Table \ref{simd} by adapting the \gls{simd} \cite{simde} library. Here, \gls{dst}-VII, \gls{dct}-II, \gls{dct}-VIII, inter-component transform and \gls{lfnst} module of \gls{tx} block of \gls{vvc} decoder was accelerated using \gls{simd} registers. \gls{tx} involves several matrix operations including matrix multiplication for the inverse transformations. These matrix operations were tackled using \gls{simd} intrinsics based on logical and mathematical operations: vand, veor, vadd, and vmul. The most benefiting modules of \gls{ep} block by \gls{simd} parallelisation are luma 8-tap filters, chroma 4-tap filters, bi-directional optical flow, decoder side motion vector refinement, and prediction refinement with optical flow. These functions contain various mathematical and clipping operations which were handled by vadd, vsub, vmin, and vmax instructions. In addition, loading and storing data in larger \gls{simd} registers helped to accelerate \gls{ep}, because the prediction information of the pixels is needed multiple times in different \gls{ep} functions. Then, the pixel prediction inside the picture of \gls{ip} block was effectively managed by storing masks, clipped, and offset value using \gls{simd} intrinsics. Further, the edge and band filter of \gls{sao} use vceq, vadd, and vsub instructions to handle mathematical operations. Finally, \gls{alf} filters are parallelized by concurrently storing filter parameters using shuffle intrinsic. Moreover, it exploits the full capacity of \gls{simd} register of 128 bit using load and store intrinsic instructions. 

\begin{table}[t] % PENDING UPDATE
\caption{Main functions optimized with \gls{simd}.}
\label{simd}
\resizebox{\columnwidth}{!}{
\centering
\begin{tabular}{cl}
\toprule
\gls{vvc} Block                                                                  & \multicolumn{1}{c}{Module}                             \\ \midrule
\multirow{3}{*}{TX}                                                        & DST-VII, DCT-II, DCT-VIII                               \\
                                                                           & Inter-component transform                              \\
                                                                           & Low-frequency non-separable transform                  \\ \midrule
\multirow{5}{*}{IP}                                                        & Luma 8-tap filters                                     \\
                                                                           & Chroma 4-tap filters                                   \\
                                                                           & Bi-directional optical flow                            \\
                                                                           & Decoder side motion vector refinement                  \\
                                                                           & Prediction refinement with optical flow                \\ \midrule
\multirow{3}{*}{EP}                                                        & DC, Planar                                             \\
                                                                           & Cross-component linear model                           \\
                                                                           & Matrix-based intra prediction module                   \\ \midrule
\multirow{4}{*}{\begin{tabular}[c]{@{}c@{}}In-loop\\ filters\end{tabular}} & Edge and band filter of SAO                            \\
                                                                           & ALF 7$\times$7 diamond shape filters for the luma component   \\
                                                                           & ALF 5$\times$5 diamond shape filters for the chroma component \\
                                                                           & Block classification of ALF                            \\ \bottomrule
\end{tabular}
}
\end{table}

\section{Experimental Results}\label{result}
In this section, the experimental setup, test bench used in this study and the experimental results obtained are presented for two open source optimized decoders \gls{vvdec} V1.3 and OpenVVC V1.0 on two \gls{egpp}-based embedded platforms.    

\subsection{Experimental Setup} 
This study focuses on low-cost mobile embedded heterogeneous platforms. Therefore, two \gls{esoc} platforms, \gls{esoc}1 \cite{xavier} and \gls{esoc}2 \cite{nano} have been used. \gls{esoc}1 processor consists of eight \gls{egpp} cores running with a maximum clock speed of 2.26 GHz and 512 embedded \gls{gpu} (E\gls{gpu}) cores running with a maximum clock speed of 1.37 GHz. In addition, \gls{esoc}1 has 8MB of L2 cache memory, 4MB of L3 cache memory and 32GB 256-Bit random access memory with 137 GB/s speed. \gls{esoc}2 has 4 \gls{egpp} cores and 128 E\gls{gpu} cores running with a maximum clock speed of 1.48GHz and 0.92GHz, respectively. Moreover, it has 2MB L2 cache memory and 4MB 64 bit random access memory with 25.6 GB/s speed. This work is only based on \gls{egpp} cores and \gls{gpu} cores could be used in future works to further speedup the decoder. In both platforms a gcc compiler version 7.5 with -O3 flag activated and Cmake version 3.16.5 have been used alongside an Ubuntu 18.04 operating system. %Lastly, all experimental results have been taken using the maximum clock speed of both platforms.

\subsection{Test video sequences}\label{seq} 
Table \ref{TB} presents the different features of the fifteen \gls{jvet} common test sequences \cite{testset} used in this study. The following sequences, grouped by resolution classes, have been encoded by the \gls{vtm}-11.0 reference software with 10-bit random access and $3\times3$ tile configuration at two \gls{qp} 27 and 37. Three \gls{hd} sequences from class E are used: FourPeople, Johnny, and KristenAndSara alongside six \gls{fhd} sequences of class B: MarketPlace, RitualDance, Cactus, BasketballDrive, BQTerrace, and ArenaOfValor. We also considered six \gls{uhd} sequences from class A1: Tango2, FoodMarket4, and Campfire and class A2: CatRobot1, DaylightRoad2, and ParkRunning3.

%Here, three class A1 sequences: Tango2 (TG2), FoodMarket4 (FM4), and Campfire (CFR); three class A2 sequences: CatRobot1 (CR1), DaylightRoad2 (DR2), and ParkRunning3 (PR3); six class B sequences: MarketPlace (MPL), RitualDance (RUD), Cactus (CCT), BasketballDrive (BBD), BQTerrace (BQT), and ArenaOfValor (AOV); and class E sequences: FourPeople (FRP), Johnny (JNY), and KristenAndSara (KAS) have been encoded with random access (RA) configuration and quantization parameters (QP) 27 and 37.

\begin{table}[t] % PENDING UPDATE
\caption{Features of the considered \gls{vvc} test sequences.}
\label{TB}
\resizebox{\columnwidth}{!}{
\centering
\begin{tabular}{clcccc}
\toprule
Class               & Sequence & Resolution   & \# Frames & Bitdepth & Framerate \\ \midrule
\multirow{3}{*}{A1} & Tango2      & 3840$\times$2160 & 294 & 10 &60 \\      %\\ \cline{2-5} 
                    & FoodMarket4      & 3840$\times$2160 & 300 & 10 &60 \\ 
                    & Campfire      & 3840$\times$2160 & 300 & 10 &30 \\      \midrule
\multirow{3}{*}{A2} & CatRobot1      & 3840$\times$2160 & 300 & 10 &60 \\      %\\ \cline{2-5} 
                    & DaylightRoad2      & 3840$\times$2160 & 300 & 10 &60 \\ 
                    & ParkRunning3      & 3840$\times$2160 & 300 & 10 &50 \\ \midrule
\multirow{6}{*}{B}  & MarketPlace      & 1920$\times$1080 & 300 & 10 &60 \\    %\\ \cline{2-5} 
                    & RitualDance      & 1920$\times$1080 & 300 & 10 &60 \\
                    & Cactus      & 1920$\times$1080 & 300 & 10 &50 \\
                    & BasketballDrive      & 1920$\times$1080 & 300 & 10 &50 \\
                    & BQTerrace      & 1920$\times$1080 & 300 & 10 &60 \\    %\\ \cline{2-5} 
                    & ArenaOfValor      & 1920$\times$1080 & 300 & 10 &60 \\ \midrule
\multirow{3}{*}{E}  & FourPeople      & 1280$\times$720  & 300 & 10 &60 \\      %\\ \cline{2-5} 
                    & Johnny      & 1280$\times$720  & 300 & 10 &60 \\ 
                    & KristenAndSara      & 1280$\times$720  & 300 & 10 &60 \\ \bottomrule
\end{tabular}
}
\end{table}

\subsection{Results and analyses}
Since this study focuses on analysing the decoding performance over embedded platforms, the average energy consumption and the maximum memory usage have been also measured. To do so,  two open source optimized \gls{vvc} decoders: \gls{vvdec} and OpenVVC over the two already mentioned platforms have been used. %Although most of the studies focus on decoding performance, energy consumption and memory usage are the key factors for hardware optimization. Especially for embedded platforms, energy consumption and memory usage are the main constrains. Therefore, the energy consumption and memory usage of OVC and VDC configurations that achieve real-time decoding are presented in this section.  

%Therefore, the comparison between VDC and OVC is demonstrated.   

%the suitability of two different VVC decoders has been tested

\subsubsection{Decoding performance}
Firstly, the decoding performance of OpenVVC has been studied for four combinations of frame-tile parallelization by taking advantage of the 8 physical cores integrated in the \gls{esoc}1 architecture:
\begin{itemize}
    \item 1-frame and 8-tile per frame in parallel (f1/t8).
    \item 2-frame and 4-tile per frame in parallel (f2/t4).
    \item 4-frame and 2-tile per frame in parallel (f4/t2).
    \item 2-frame and 8-tile per frame in parallel (f2/t8).
\end{itemize}

This study has been performed to do a fair comparison between the best configuration of OpenVVC and VVdeC.

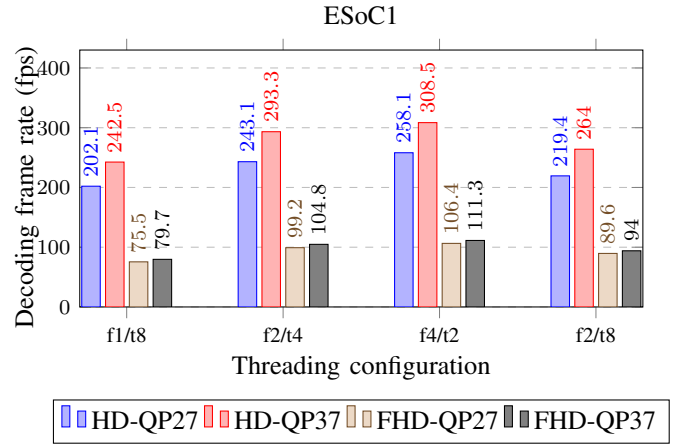
\begin{figure}[hbt!]
\centering
%\begin{tabular}{W{c}{0.49\textwidth}|W{c}{0.49\textwidth}}
\begin{tikzpicture}  
\begin{axis}  
[   title={ESoC1},
    width=0.5\textwidth,height=50mm,
    ymin=0, ymax=430, 
    ybar, % configures ‘bar shift’
    ymajorgrids=true,
    grid style=dashed,
    xtick=data,
    ylabel={Decoding frame rate (\gls{fps})},ylabel style={xshift=-60pt,yshift=-15pt,anchor=west},
    xlabel={Threading configuration},%anchor=east,
    tick label style={font=\footnotesize},
    bar width=7pt,% height=54mm,
    %legend style={at={(0.195,0.95)}, anchor=north, {nodes={scale=0.6, transform shape}}},
    %ylabel={\% load per block}, % there should be no line gap between the rows here. Otherwise, latex will show an error.  
    legend style={at={(0.5,-0.36)},
	anchor=north,legend columns=-1},
	%ybar interval=0.3,
    symbolic x coords={f1/t8, f2/t4, f4/t2, f2/t8},  
    nodes near coords,  
    every node near coord/.append style={rotate=90, xshift=0cm, yshift=0cm, anchor=west,font=\footnotesize},
    nodes near coords align={vertical},  
    ]  
\addplot coordinates {(f1/t8, 202.1) (f2/t4, 243.1) (f4/t2, 258.1) (f2/t8, 219.4)};
\addplot coordinates {(f1/t8, 242.5) (f2/t4, 293.3) (f4/t2, 308.5) (f2/t8, 264.0)};
\addplot coordinates {(f1/t8, 75.5) (f2/t4, 99.2) (f4/t2, 106.4) (f2/t8, 89.6)};
\addplot coordinates {(f1/t8, 79.7) (f2/t4, 104.8) (f4/t2, 111.3) (f2/t8, 94.0)};

\legend{HD-QP27, HD-QP37, FHD-QP27, FHD-QP37} 
\end{axis}  
\end{tikzpicture}
%\end{tabular}

%\caption{} %A caption inside a tabular section may cause errors
\caption{Average decoding performance (\gls{fps}) of OpenVVC for QP 27 and 37 sequences on the ESoC1 platform.}
\label{xfps}
\end{figure}

Fig.\ref{xfps} shows the average decoding performance in frames per second (\gls{fps}) for HD and FHD test sequences with QP27 and QP37 on ESoC1 using OpenVVC decoder.

It can be seen from Fig.\ref{xfps} that the least performing configuration is f1/t8 and the best performing configuration is f4/t2 for all \gls{qp}s, \gls{hd}, and \gls{fhd} sequences. f4/t2 configuration has achieved in average $\times 1.4$ and $\times 1.3$ \gls{fps} compared to the f1/t8 configuration for \gls{fhd} and \gls{hd} sequences, respectively. %In addition, the f4/t2 configuration has obtained average 6\% more FPS than f2/t4 and average 17\% more FPS than the f2/t8 configuration for all qualities of the sequences.   
These results mainly illustrate the gain brought considering both frame and tile parallelism compared to only frame parallelism which is constrained by the inter coding dependency.

\begin{figure}[hbt!]
\centering
%\begin{tabular}{W{c}{0.49\textwidth}|W{c}{0.49\textwidth}}
\begin{tikzpicture}  
\begin{axis}  
[   title={ESoC2},
    width=0.5\textwidth,height=50mm,
    ymin=0, ymax=100,
    ybar, % configures ‘bar shift’
    ymajorgrids=true,
    grid style=dashed,
    xtick=data,
    ylabel={Decoding frame rate (\gls{fps})},ylabel style={xshift=-60pt,yshift=-15pt,anchor=west},
    xlabel={Thread configuration},%anchor=east,
    tick label style={font=\footnotesize},
    %bar width=7pt,% height=54mm,
    %legend style={at={(0.28,0.95)}, anchor=north, {nodes={scale=0.8, transform shape}}},
    %ylabel={\% load per block}, % there should be no line gap between the rows here. Otherwise, latex will show an error.  
    legend style={at={(0.5,-0.36)},
	anchor=north,legend columns=-1},
    symbolic x coords={f1/t4, f2/t2, f2/t4},  
    nodes near coords,  
    every node near coord/.append style={rotate=90, xshift=0cm, yshift=0cm, anchor=west,font=\footnotesize},
    nodes near coords align={vertical},  
    ]  
\addplot coordinates {(f1/t4, 55.6) (f2/t2, 62.6) (f2/t4, 59.7)};
\addplot coordinates {(f1/t4, 65.2) (f2/t2, 74.5) (f2/t4, 70.6)};

\legend{HD-QP27, HD-QP37} 
\end{axis}  
\end{tikzpicture}

\caption{Average decoding performance (\gls{fps}) of the OpenVVC decoder for QPs 27 and 37 sequences on the ESoC2 platform.}
\label{nfps}
\end{figure}
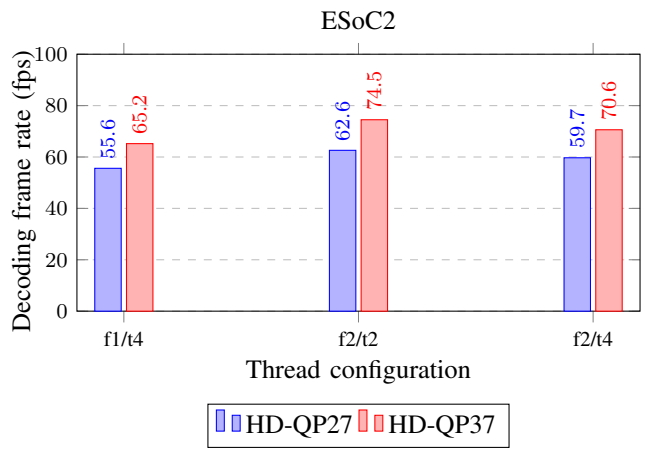

Fig.\ref{nfps} presents the average decoding performance in \gls{fps} obtained over \gls{esoc}2 for \gls{hd} sequences using both \gls{qp}s. \gls{esoc}2 has four physical cores. For this reason, only three combinations of frame-title parallelization have been studied: 
\begin{itemize}
    \item 1-frame and 4-tile per frame in parallel (f1/t4).
    \item 2-frame and 2-tile per frame in parallel (f2/t2).
    \item 2-frame and 4-tile per frame in parallel (f2/t4).
\end{itemize}

In this case, and as shown in Fig.\ref{nfps}, the f2/t2 configuration has achieved 62.6 \gls{fps} for \gls{qp}27 and 74.5 \gls{fps} for \gls{qp}37, which is higher in average by  8.1 \gls{fps} and 3.4 \gls{fps} than f1/t4 and f2/t4 configurations for \gls{hd} sequences, respectively.  

The decoding performance and speedup of \gls{vvdec} and f4/t2 configuration of OpenVVC decoders for \gls{qp}s 27 and 37 over \gls{esoc}1 are presented in Table~\ref{xal} for all considered video sequences. It can be seen that the obtained decoding speed by both \gls{vvdec} and OpenVVC decoders are close to real-time for \gls{uhd} sequences and achieve real-time for all \gls{fhd} and \gls{hd} sequences over \gls{esoc}1 using 8 cores. Therefore, the experiment for \gls{fhd} and \gls{hd} sequences are presented in the following part of this study that achieves real-time over \gls{esoc}1. Moreover, it can be noticed that \gls{vvdec} has obtained slightly better \gls{fps} than OpenVVC on single core configuration for all sequences at different quantification parameters QPs. However, speedup shows that OpenVVC has provided better parallelism compared to \gls{vvdec} when the number of threads has increased, which compensates the first limitation.

\begin{table*}[t]
\caption{Decoding performance (\gls{fps}) for the considered test sequences at QP27 (top) and QP37 (bottom) on the ESoC1 platform with 1, 2, 4, 6 and 8 cores.}
\label{xal}
%\resizebox{\columnwidth}{!}{
\centering
%\begin{adjustbox}{max width=0.48\textwidth}%\textwidth
\begin{tabular}{l|ccccc|ccccc}
\toprule
Seq.:QP    & \multicolumn{5}{c|}{VVdeC (\gls{fps})/speedup}      & \multicolumn{5}{c}{OpenVVC (\gls{fps})/speedup}       \\ \midrule
\# cores & 1    & 2     & 4     & 6     & 8     & 1    & 2     & 4     & 6     & 8     \\ \midrule
Tango2:27     & 3.3/1.0  & 6.3/1.9   & 12.4/3.8  & 18.1/5.6  & 22.8/7.0  & 3.1/1.0  & 6.1/2.0   & 12.1/3.9  & 17.9/5.8  & 22.7/7.4  \\
FoodMarket4:27     & 3.3/1.0  & 5.9/1.8   & 12.4/3.8  & 18.2/5.6  & 23.0/7.0  & 3.0/1.0  & 6.1/2.0   & 12.0/4.0  & 17.6/5.8  & 22.6/7.5  \\
Campfire:27     & 3.7/1.0  & 6.8/1.9 & 13.6/3.7  & 19.5/5.3  & 24.0/6.5  & 3.5/1.0 & 7.5/2.1   & 14.7/4.2  & 21.7/6.2  & 27.3/7.8  \\ 
{\bf Average}      &3.4/1.0  &6.4/1.9   &12.8/3.8   &18.6/5.5  &23.3/6.8  &3.2/1.0 &6.6/2.0 &12.9/4.0 &19.0/6.0 &24.2/7.6  \\ \midrule
CatRobot1:27     & 3.3/1.0  & 6.4/1.9   & 12.7/3.8  & 18.5/5.5  & 23.3/7.0  & 3.2/1.0  & 6.3/2.0 & 12.4/3.9  & 18.1/5.7  & 22.9/7.2  \\
DaylightRoad2:27     & 3.1/1.0  & 6.0/1.9   & 11.7/3.8  & 17.2/5.5  & 21.5/7.0  & 2.8/1.0  & 5.7/2.0   & 11.2/4.0  & 16.4/5.9  & 20.9/7.5  \\
ParkRunning3:27     & 2.4/1.0  & 4.3/1.8   & 8.5/3.6   & 12.4/5.2  & 15.7/6.7  & 2.1/1.0  & 4.1/2.0   & 8.1/3.9   & 11.9/5.7  & 15.1/7.3  \\
{\bf Average}       &2.9/1.0  &5.6/1.9   &11.0/3.7   &16.0/5.5  &20.2/6.9  &2.7/1.0 &5.3/2.0 &10.5/3.9 &15.5/5.8 &19.7/7.3  \\ \midrule
MarketPlace:27     & 12.5/1.0 & 23.8/1.9  & 46.0/3.7  & 66.5/5.3  & 82.7/6.6  & 11.1/1.0 & 21.8/2.0  & 43.2/3.9  & 63.3/5.7  & 79.6/7.2  \\
RitualDance:27     & 13.9/1.0 & 26.3/1.9  & 51.3/3.7  & 73.2/5.2  & 91.0/6.5  & 13.0/1.0 & 25.2/1.9  & 49.8/3.8  & 73.7/5.7  & 90.4/7.0  \\
Cactus:27     & 16.6/1.0 & 30.7/1.8  & 59.0/3.6  & 85.1/5.1  & 103.0/6.2 & 15.4/1.0 & 29.6/1.9  & 57.9/3.8  & 82.6/5.4  & 94.3/6.1  \\
BasketballDrive:27   & 12.2/1.0 & 22.5/1.8  & 43.7/3.6  & 63.5/5.2  & 77.5/6.4  & 11.1/1.0 & 21.5/1.9  & 42.8/3.9  & 63.2/5.7  & 79.6/7.2  \\
BQTerrace:27     & 12.9/1.0 & 24.3/1.9  & 47.4/3.7  & 68.7/5.3  & 82.6/6.4  & 11.7/1.0 & 23.1/2.0  & 45.7/3.9  & 67.4/5.8  & 85.2/7.3  \\
ArenaOfValor:27     & 15.5/1.0 & 28.5/1.8  & 55.0/3.6  & 79.2/5.1  & 96.8/6.3  & 14.0/1.0 & 27.3/1.9  & 53.8/3.8  & 78.9/5.6  & 96.8/6.9  \\
{\bf Average}      &13.9/1.0  &26.0/1.9   &50.4/3.6   &72.7/5.2  &88.9/6.4  &12.7/1.0  &24.7/1.9 &48.9/3.8 &71.5/5.6 &87.6/6.9  \\ \midrule
FourPeople:27     & 52.2/1.0 & 98.7/1.9  & 185.5/3.6 & 257.1/4.9 & 271.0/5.2 & 45.8/1.0 & 88.8/1.9  & 175.4/3.8 & 252.1/5.5 & 285.7/6.2 \\
Johnny:27     & 54.3/1.0 & 104.4/1.9 & 196.7/3.6 & 272.2/5.0 & 288.2/5.3 & 48.9/1.0 & 92.6/1.9  & 180.7/3.7 & 238.1/4.9 & 252.1/5.2 \\
KristenAndSara:27  & 52.0/1.0 & 97.3/1.9  & 184.0/3.5 & 257.5/5.0 & 294.4/5.7 & 47.6/1.0 & 89.6/1.9  & 177.5/3.7 & 250.0/5.3 & 272.7/5.7 \\
{\bf Average}     &52.8/1.0  &100.1/1.9   &188.8/3.6 &262.3/5.0  &284.5/5.4  &47.5/1.0  &90.3/1.9 &177.9/3.7 &246.7/5.2 &270.2/5.7  \\ \midrule \midrule
Tango2:37     & 4.1/1.0  & 8.2/2.0   & 16.0/3.9  & 23.2/5.6  & 29.7/7.2  & 4.1/1.0  & 8.2/2.0   & 15.1/3.7  & 23.7/5.8  & 30.0/7.3  \\
FoodMarket4:37     & 4.0/1.0  & 7.9/2.0   & 15.8/3.9  & 22.9/5.7  & 29.2/7.2  & 3.9/1.0  & 7.8/2.0   & 15.3/3.9  & 22.5/5.7  & 28.2/7.2  \\
Campfire:37     & 4.8/1.0  & 9.0/1.9   & 17.8/3.7  & 25.8/5.4  & 31.9/6.7  & 4.7/1.0  & 9.7/2.0   & 19.0/4.0  & 28.1/5.9  & 34.3/7.2  \\
{\bf Average}      &4.3/1.0  &8.4/1.9   &16.5/3.8   &24.0/5.6  &30.3/7.0  &4.3/1.0  &8.6/2.0   &16.5/3.9   &24.8/5.8  &30.8/7.2  \\ \midrule
CatRobot1:37     & 4.1/1.0  & 8.2/2.0   & 16.1/3.9  & 23.3/5.7  & 29.4/7.1  & 4.2/1.0  & 8.3/2.0   & 16.2/3.9  & 23.8/5.7  & 30.0/7.2  \\
DaylightRoad2:37     & 4.1/1.0  & 8.0/2.0 & 16.0/3.9  & 23.4/5.7  & 29.6/7.2  & 3.8/1.0  & 7.8/2.0   & 15.4/4.0  & 22.6/5.9  & 28.5/7.4  \\
ParkRunning3:37     & 3.0/1.0  & 5.8/1.9  & 11.5/3.8  & 16.9/5.6  & 21.3/7.1  & 2.8/1.0  & 5.5/2.0   & 10.8/3.9  & 16.1/5.8  & 20.6/7.4  \\
{\bf Average}      &3.8/1.0  &7.3/2.0   &14.5/3.9   &21.2/5.6  &26.8/7.1  &3.6/1.0  &7.2/2.0   &14.2/3.9   &20.8/5.8  &26.4/7.3  \\ \midrule
MarketPlace:37     & 16.9/1.0 & 33.2/2.0  & 64.3/3.8  & 92.6/5.5  & 109.9/6.5 & 16.0/1.0 & 31.0/1.9  & 61.2/3.8  & 89.8/5.6  & 107.1/6.7 \\
RitualDance:37     & 18.0/1.0 & 34.9/1.9  & 67.2/3.7  & 97.0/5.4  & 119.8/6.6 & 16.6/1.0 & 32.4/1.9  & 64.7/3.9  & 95.8/5.8  & 117.2/7.0 \\
Cactus:37     & 21.7/1.0 & 40.3/1.9  & 77.3/3.6  & 111.0/5.1 & 133.0/6.1 & 20.5/1.0 & 39.5/1.9  & 76.9/3.8  & 111.1/5.4 & 124.0/6.0 \\
BasketballDrive:37     & 14.7/1.0 & 28.1/1.9  & 55.0/3.7  & 79.6/5.4  & 98.2/6.7  & 13.7/1.0 & 26.6/1.9  & 52.9/3.9  & 77.5/5.6  & 97.1/7.1  \\
BQTerrace:37     & 15.8/1.0 & 30.6/1.9  & 59.7/3.8  & 87.4/5.5  & 104.1/6.6 & 15.5/1.0 & 30.3/2.0  & 60.1/3.9  & 88.2/5.7  & 109.1/7.0 \\
ArenaOfValor:37     & 19.9/1.0 & 38.5/1.9 & 74.4/3.7  & 106.6/5.4 & 128.7/6.5 & 19.1/1.0 & 37.9/2.0  & 74.6/3.9  & 108.3/5.7 & 130.4/6.8 \\ 
{\bf Average}     &17.8/1.0  &34.3/1.9   &66.3/3.7   &95.7/5.4  &115.6/6.5  &16.9/1.0  &33.0/1.9   &65.1/3.8   &95.1/5.6  &114.2/6.7  \\ 
\midrule
FourPeople:37     & 59.2/1.0 & 116.7/2.0 & 219.1/3.7 & 299.4/5.1 & 314.5/5.3 & 55.5/1.0 & 108.3/2.0 & 211.3/3.8 & 306.1/5.5 & 344.8/6.2 \\
Johnny:37     & 64.2/1.0 & 123.2/1.9 & 231.5/3.6 & 319.5/5.0 & 355.0/5.5 & 57.8/1.0 & 113.2/2.0 & 219.0/3.8 & 297.0/5.1 & 319.1/5.5 \\
KristenAndSara:37     & 60.0/1.0 & 115.4/1.9 & 217.5/3.6 & 302.1/5.0 & 338.2/5.6 & 52.9/1.0 & 103.8/2.0 & 204.1/3.9 & 294.1/5.6 & 322.6/6.1 \\ 
{\bf Average}     &61.1/1.0  &118.4/1.9   &222.7/3.6   &307.0/5.0  &335.9/5.5  &55.4/1.0  &108.4/2.0   &211.4/3.8   &299.1/5.4  &328.9/5.9  \\  \bottomrule
\end{tabular}
%}
%\end{adjustbox}
\end{table*}

In Table \ref{nol}, the decoding performance of \gls{vvdec} and OpenVVC (f2/t2 configuration) decoders for all considered \gls{fhd} and \gls{hd} sequences with \gls{qp}s 27 and 37 over \gls{esoc}2 platform is shown. Here, both \gls{vvdec} and OpenVVC decoders have achieved real-time for \gls{hd} sequences over \gls{esoc}2 using 4 cores. Therefore, in the upcoming part of this study, the results are presented for \gls{hd} sequences over \gls{esoc}2.

\begin{table}[t] 
\caption{Decoding performance (in \gls{fps}) for the considered HD and FHD test sequences at QP27 (top) and QP37 (bottom) on the ESoC2 platform with 1, 2, 3 and 4 cores.}
\label{nol}
\centering
\resizebox{\columnwidth}{!}{
\begin{tabular}{lcccc|cccc}
\toprule
Seq.:QP    & \multicolumn{4}{c|}{VVdeC (\gls{fps})} & \multicolumn{4}{c}{OpenVVC (\gls{fps})} \\ \midrule
\# cores & 1      & 2      & 3     & 4     & 1      & 2     & 3     & 4     \\ \midrule
MarketPlace:27     & 4.6    & 8.8    & 12.9  & 16.5  & 4.3    & 8.2   & 12.0  & 15.5  \\
RitualDance:27     & 5.4    & 10.0   & 14.6  & 18.7  & 5.0    & 9.7   & 14.1  & 18.2  \\
Cactus:27     & 6.4    & 11.6   & 17.0  & 21.5  & 5.8    & 11.3  & 16.4  & 21.0  \\
BasketballDrive:27     & 4.6    & 8.6    & 12.5  & 16.2  & 4.3    & 8.5   & 12.4  & 16.1  \\
BQTerrace:27     & 4.8    & 9.0    & 13.2  & 17.1  & 4.6    & 8.9   & 13.0  & 16.7  \\
ArenaOfValor:27     & 6.0    & 10.9   & 15.9  & 20.1  & 5.5    & 10.7  & 15.7  & 20.0  \\
{\bf Average}      &5.3  &9.8   &14.3  &18.4  &4.9  &9.6  &13.9   &17.9       \\ \midrule
FourPeople:27     & 21.3   & 39.0   & 56.4  & 69.4  & 18.3   & 35.4  & 50.9  & 64.4  \\
Johnny:27     & 21.9   & 40.4   & 57.9  & 71.4  & 18.1   & 34.7  & 50.5  & 63.2  \\
KristenAndSara:27     & 20.7   & 38.1   & 55.1  & 68.3  & 18.0   & 34.8  & 50.3  & 64.0  \\ 
{\bf Average}      &21.3  &39.2   &56.5   &69.7  &18.1  &34.9  &50.6 &63.8     \\ \midrule \midrule
MarketPlace:37     & 6.3    & 12.1   & 17.6  & 22.5  & 5.9    & 11.5  & 16.9  & 21.7  \\
RitualDance:37     & 7.0    & 13.1   & 17.7  & 24.5  & 6.4    & 12.4  & 18.0  & 23.1  \\
Cactus:37     & 8.2    & 15.4   & 22.5  & 28.4  & 7.7    & 14.9  & 21.4  & 27.6  \\
BasketballDrive:37     & 5.6    & 10.7   & 15.7  & 20.3  & 5.2    & 10.3  & 14.9  & 19.2  \\
BQTerrace:37     & 5.9    & 11.4   & 16.6  & 21.4  & 5.8    & 11.4  & 16.6  & 21.4  \\
ArenaOfValor:37     & 7.8    & 14.7   & 21.4  & 26.1  & 7.5    & 14.5  & 21.0  & 26.7  \\
{\bf Average}      &6.8  &12.9   &18.6   &23.9  &6.4  &12.5  &18.1   &23.3    \\ \midrule
FourPeople:37     & 24.8   & 46.2   & 65.7  & 80.9  & 21.8   & 42.1  & 61.0  & 77.3  \\
Johnny:37     & 25.6   & 48.1   & 68.6  & 83.6  & 21.9   & 42.3  & 60.9  & 76.9  \\
KristenAndSara:37     & 23.8   & 44.5   & 63.8  & 77.9  & 20.6   & 39.6  & 57.0  & 72.6  \\ 
{\bf Average}      &24.7  &46.2   &66.0   &80.8  &21.4  &41.3  &59.6 &75.6     \\ \bottomrule
\end{tabular}
}
\end{table}

 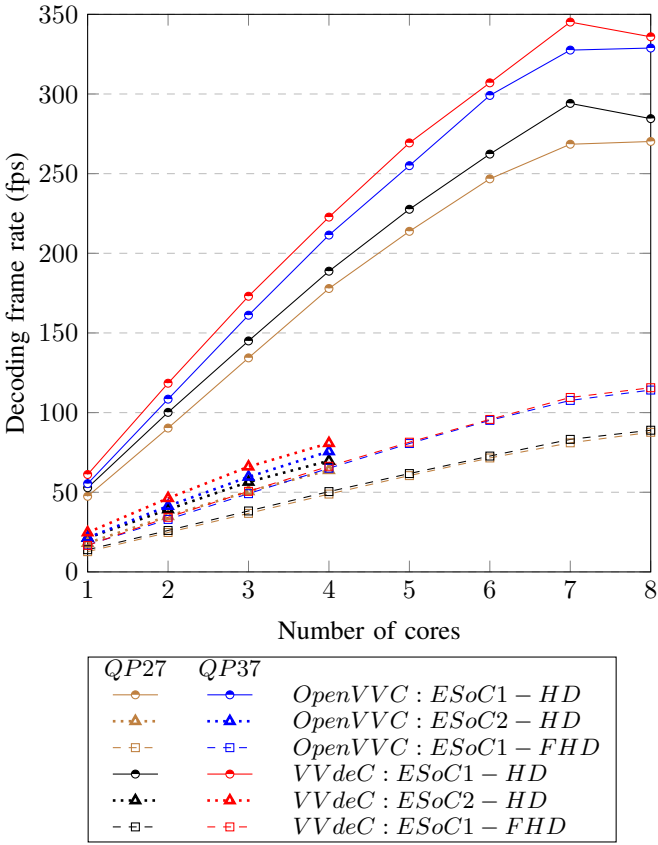
\begin{figure}[hbt!]

\begin{tikzpicture}
\begin{axis}[
    width=0.5\textwidth,height=90mm,
    ymin=0, ymax=350,
    xlabel={Number of cores},
    ylabel={Decoding frame rate (\gls{fps})},
    y label style={at={(0.04,0.5)}},
    xmin=1, xmax=8,
    %ymin=0, %ymax=350,
    xtick={1,2,3,4,5,6,7,8},
    ytick={0,50,100,150,200,250,300,350},
    %legend pos=north west,
    legend style={at={(0.24,0.98)}, anchor=north, {nodes={scale=0.7, transform shape}}},
    ymajorgrids=true,
    grid style=dashed,
]

\addplot [color=brown, mark=halfcircle*, mark options={solid}, mark size = 1.5pt]coordinates {(1,47.45)(2,90.30)(3,134.35)(4,177.89)(5,213.79)(6,246.73)(7,268.46)(8,270.18)}; \label{pgfplots:c1r1}
\addplot [line width=0.35mm, color=brown, dotted, nodes near coords align={horizontal}, mark=triangle, mark options={solid}, mark size = 2pt]coordinates {(1,18.12)(2,34.94)(3,50.56)(4,63.83)}; \label{pgfplots:c1r2}
\addplot [color=brown, dashed, nodes near coords align={horizontal}, mark=square, mark options={solid}, mark size = 1.5pt]coordinates {(1,12.72)(2,24.74)(3,36.75)(4,48.86)(5,60.45)(6,71.53)(7,81.04)(8,87.64)}; \label{pgfplots:c1r3}

\addplot [color=black, mark=halfcircle*, mark options={solid}, mark size = 1.5pt]coordinates {(1,52.81)(2,100.15)(3,144.99)(4,188.77)(5,227.63)(6,262.27)(7,294.09)(8,284.53)}; \label{pgfplots:c2r1}
\addplot [line width=0.35mm, color=black, dotted, nodes near coords align={horizontal}, mark=triangle, mark options={solid}, mark size = 2pt]coordinates {(1,21.31)(2,39.18)(3,56.48)(4,69.74)}; \label{pgfplots:c2r2}
\addplot [color=black, dashed, nodes near coords align={horizontal}, mark=square, mark options={solid}, mark size = 1.5pt]coordinates {(1,13.94)(2,26.01)(3,38.26)(4,50.42)(5,61.73)(6,72.72)(7,83.19)(8,88.94)}; \label{pgfplots:c2r3}

\addplot [color=blue, mark=halfcircle*, mark options={solid}, mark size = 1.5pt]coordinates {(1,55.39)(2,108.44)(3,161.13)(4,211.44)(5,255.01)(6,299.09)(7,327.55)(8,328.85)}; \label{pgfplots:c3r1}
\addplot  [line width=0.35mm, color=blue, dotted, nodes near coords align={horizontal}, mark=triangle, mark options={solid}, mark size = 2pt]coordinates {(1,21.43)(2,41.34)(3,59.62)(4,75.63)}; \label{pgfplots:c3r2}
\addplot [color=blue, dashed, nodes near coords align={horizontal}, mark=square, mark options={solid}, mark size = 1.5pt]coordinates {(1,16.92)(2,32.95)(3,49.12)(4,65.08)(5,80.59)(6,95.14)(7,107.65)(8,114.15)}; \label{pgfplots:c3r3}

\addplot [color=red, mark=halfcircle*, mark options={solid}, mark size = 1.5pt]coordinates {(1,61.14)(2,118.44)(3,173.04)(4,222.72)(5,269.30)(6,307.00)(7,345.15)(8,335.90)}; \label{pgfplots:c4r1}
\addplot [line width=0.35mm, color=red, dotted, nodes near coords align={horizontal}, mark=triangle, mark options={solid}, mark size = 2pt]coordinates {(1,24.72)(2,46.25)(3,66.04)(4,80.78)}; \label{pgfplots:c4r2}
\addplot [color=red, dashed, nodes near coords align={horizontal}, mark=square, mark options={solid}, mark size = 1.5pt] coordinates {(1,16.66)(2,34.28)(3,50.57)(4,66.32)(5,81.43)(6,95.71)(7,109.50)(8,115.62)}; \label{pgfplots:c4r3}
%     \legend{OpenVVC:ESoC1-HD, OpenVVC:ESoC2-HD, OpenVVC:ESoC1-FHD, VVdeC:ESoC1-HD, VVdeC:ESoC2-HD, VVdeC:ESoC1-FHD}
    %\legend{OpenVVC:ESoC1-HD, OpenVVC:ESoC2-HD, OpenVVC:ESoC1-FHD, VVdeC:ESoC1-HD, VVdeC:ESoC2-HD, VVdeC:ESoC1-FHD}
\end{axis}

     \node[draw,fill=white,inner sep=0pt, yshift=-3.6cm, above right]
     {\small
    \begin{tabular}{ccl}
    $QP27$ & $QP37$ \\
    \ref{pgfplots:c1r1} & \ref{pgfplots:c3r1} & $OpenVVC:ESoC1-HD$\\
    \ref{pgfplots:c1r2} & \ref{pgfplots:c3r2} & $OpenVVC:ESoC2-HD$\\
    \ref{pgfplots:c1r3} & \ref{pgfplots:c3r3} & $OpenVVC:ESoC1-FHD$\\
    \ref{pgfplots:c2r1} & \ref{pgfplots:c4r1} & $VVdeC:ESoC1-HD$\\
    \ref{pgfplots:c2r2} & \ref{pgfplots:c4r2} & $VVdeC:ESoC2-HD$\\
    \ref{pgfplots:c2r3} & \ref{pgfplots:c4r3} & $VVdeC:ESoC1-FHD$
    \end{tabular}};

\end{tikzpicture}
\caption{Average decoding performance (in \gls{fps}) of OpenVVC, in brown QP 27 and blue QP 37, and VVdeC, in black QP27 and red QP37, for 1 to 8 number of cores.}
\label{nxcore}
\end{figure}

The average decoding performance with respect to the number of cores is presented in the Fig. \ref{nxcore}. Here, the decoding frame rates have been recorded for OpenVVC and \gls{vvdec} decoders over \gls{esoc}1 and \gls{esoc}2. For both \gls{qp}s, the average results in \gls{fps} of OpenVVC and \gls{vvdec} are similar for one to four cores over \gls{esoc}2. Moreover, \gls{vvdec} has achieved up to $\times 1.08$ \gls{fps} with respect to OpenVVC on \gls{esoc}1 and it has reached the saturation point with 7 cores for \gls{hd} sequences. However, for \gls{fhd} sequences, the performance results of OpenVVC and \gls{vvdec} are comparable on \gls{esoc}1. The performance results follow the same pattern for both considered \gls{qp}s.     

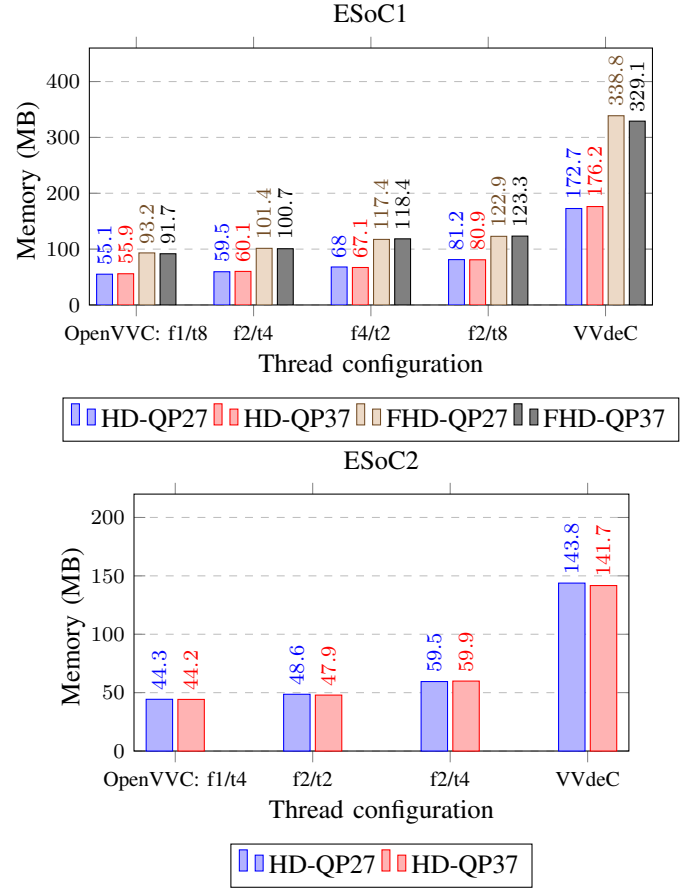
\begin{figure}[t]
\centering
%\begin{tabular}{W{c}{0.49\textwidth}|W{c}{0.49\textwidth}}
\begin{tikzpicture}  
\begin{axis}  
[   title={ESoC1},
    width=0.5\textwidth,height=50mm,
    ymin=0,
    ybar, % configures ‘bar shift’
    ymajorgrids=true,
    grid style=dashed,
    xtick=data, ymax=460,
    ytick={0,100,200,300,400},
    tick label style={font=\footnotesize},
    xlabel={Thread configuration},
    ylabel={Memory (MB)},ylabel style={xshift=-35pt,yshift=-12pt,anchor=west},
    bar width=6pt,% height=54mm,
    legend style={at={(0.5,-0.36)}, anchor=north,legend columns=-1},
    %ylabel={\% load per block}, % there should be no line gap between the rows here. Otherwise, latex will show an error.  
    symbolic x coords={OpenVVC: f1/t8, f2/t4, f4/t2, f2/t8, VVdeC},  
    nodes near coords,  
    every node near coord/.append style={rotate=90, xshift=0cm, yshift=0cm, anchor=west,font=\footnotesize},
    nodes near coords align={vertical},  
    ]  
\addplot coordinates {(OpenVVC: f1/t8, 55.1) (f2/t4, 59.5) (f4/t2, 68.0) (f2/t8, 81.2) (VVdeC, 172.7)};
\addplot coordinates {(OpenVVC: f1/t8, 55.9) (f2/t4, 60.1) (f4/t2, 67.1) (f2/t8, 80.9) (VVdeC, 176.2)};
\addplot coordinates {(OpenVVC: f1/t8, 93.2) (f2/t4, 101.4) (f4/t2, 117.4) (f2/t8, 122.9) (VVdeC, 338.8)};
\addplot coordinates {(OpenVVC: f1/t8, 91.7) (f2/t4, 100.7) (f4/t2, 118.4) (f2/t8, 123.3) (VVdeC, 329.1)};

\legend{HD-QP27, HD-QP37, FHD-QP27, FHD-QP37} 
\end{axis}  
\end{tikzpicture}
%\end{tabular}

%\caption{} %A caption inside a tabular section may cause errors
\begin{tikzpicture}  
\begin{axis}  
[   title={ESoC2},
    width=0.45\textwidth,height=50mm,
    ymin=0, ymax=220,
    ybar, % configures ‘bar shift’
    ymajorgrids=true,
    grid style=dashed,
    xtick=data,
    tick label style={font=\footnotesize},
    xlabel={Thread configuration},
    ylabel={Memory (MB)},ylabel style={xshift=-35pt,yshift=-12pt,anchor=west},
    legend style={at={(0.5,-0.36)}, anchor=north,legend columns=-1},
    %ylabel={\% load per block}, % there should be no line gap between the rows here. Otherwise, latex will show an error.  
    symbolic x coords={OpenVVC: f1/t4, f2/t2, f2/t4, VVdeC},  
    nodes near coords,  
    every node near coord/.append style={rotate=90, xshift=0cm, yshift=0cm, anchor=west,font=\footnotesize},
    nodes near coords align={vertical},  
    ]  
\addplot coordinates {(OpenVVC: f1/t4, 44.3) (f2/t2, 48.6) (f2/t4, 59.5) (VVdeC, 143.8)};%10^3
\addplot coordinates {(OpenVVC: f1/t4, 44.2) (f2/t2, 47.9) (f2/t4, 59.9) (VVdeC, 141.7)};

\legend{HD-QP27, HD-QP37}   
\end{axis}  
\end{tikzpicture}
\caption{Average maximum memory (in MB) used for QPs 27 and 37 sequences over ESoC1 (top) and ESoC2 (bottom).}
\label{xnmem}
\end{figure}

\subsubsection{Memory usage}
Memory usage is one of the most limiting factors and a likely bottleneck for video decoding over resource-constrained embedded hardware. This part of the study presents the maximum memory (in MB) consumed by OpenVVC and \gls{vvdec} over \gls{esoc}1 and \gls{esoc}2 for the \gls{fhd}/\gls{hd} sequences with two \gls{qp}s (27, 37). In Fig. \ref{xnmem}, the average maximum memory usage for different OpenVVC configurations and \gls{vvdec} is shown. Here, for both \gls{fhd} and \gls{hd} sequences, f1/t8 and f2/t8 configurations have used the least and the most memory, respectively. These behavior is expected that with the increase of the number of frames decoded in parallel, the memory usage increases since the large part memory of the decoder is related to the decoded frame. However, \gls{vvdec} requires in average $\times2.1$ more memory for \gls{hd} sequences and $\times2.7$ more memory for \gls{fhd} sequences than the OpenVVC f2/t8 configuration over \gls{esoc}1. In addition, the scenario is the same over \gls{esoc}2 platform where f2/t4 configuration requires in average a maximum memory of 59.7MB for \gls{hd} sequences. Moreover, \gls{vvdec} requires in average $\times2.4$ more memory for \gls{hd} sequences than the OpenVVC f2/t4 configuration on \gls{esoc}2 platform. It can be concluded from the results that OpenVVC requires notably less memory compared to \gls{vvdec}. Therefore, OpenVVC provides a great advantage and thus suitable for resource-constrained embedded platforms.

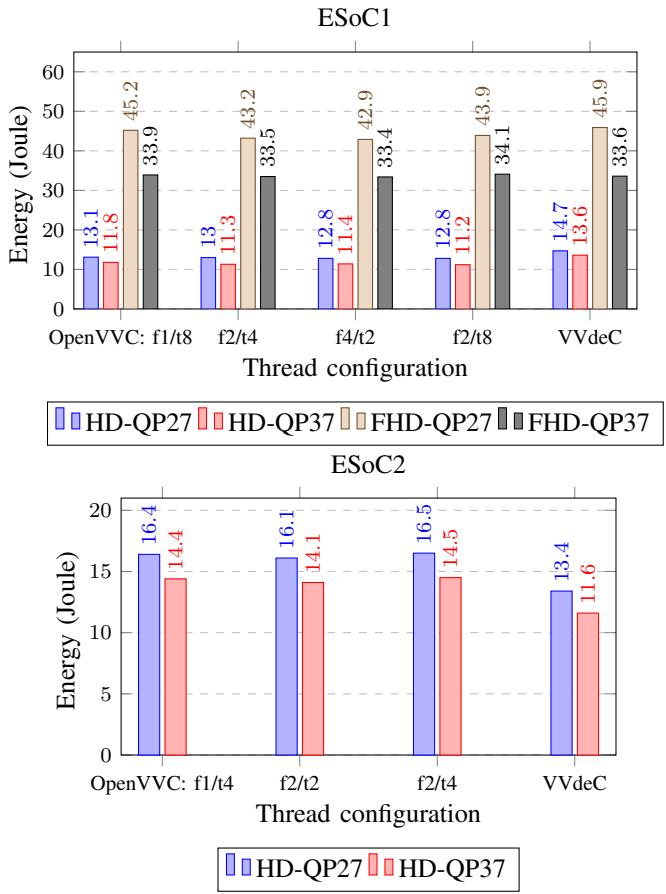
\begin{figure}[t]
\centering
%\begin{tabular}{W{c}{0.49\textwidth}|W{c}{0.49\textwidth}}
\begin{tikzpicture}  
\begin{axis}  
[   title={ESoC1},
    width=0.5\textwidth,height=50mm,
    ymin=0, ymax=65,
    ybar, % configures ‘bar shift’
    ymajorgrids=true,
    grid style=dashed,
    xtick=data, 
    ytick={0,10,20,30,40,50,60,70},
    tick label style={font=\footnotesize},
    xlabel={Thread configuration},
    ylabel={Energy (Joule)},%ylabel style={rotate=-90},
    ylabel style={xshift=-35pt,yshift=-15pt,anchor=west},
    bar width=5.5pt,% height=54mm,
    %legend style={at={(0.12,0.95)}, anchor=north, {nodes={scale=0.8, transform shape}}},
    %ylabel={\% load per block}, % there should be no line gap between the rows here. Otherwise, latex will show an error.  
    legend style={at={(0.5,-0.36)},
	anchor=north,legend columns=-1},
    symbolic x coords={OpenVVC: f1/t8, f2/t4, f4/t2, f2/t8, VVdeC},  
    nodes near coords,  
    every node near coord/.append style={rotate=90, xshift=0cm, yshift=0cm, anchor=west,font=\footnotesize},
    nodes near coords align={vertical},  
    ]  
\addplot coordinates {(OpenVVC: f1/t8, 13.1) (f2/t4, 13.0) (f4/t2, 12.8) (f2/t8, 12.8) (VVdeC, 14.7)};
\addplot coordinates {(OpenVVC: f1/t8, 11.8) (f2/t4, 11.3) (f4/t2, 11.4) (f2/t8, 11.2) (VVdeC, 13.6)};
\addplot coordinates {(OpenVVC: f1/t8, 45.2) (f2/t4, 43.2) (f4/t2, 42.9) (f2/t8, 43.9) (VVdeC, 45.9)};
\addplot coordinates {(OpenVVC: f1/t8, 33.9) (f2/t4, 33.5) (f4/t2, 33.4) (f2/t8, 34.1) (VVdeC, 33.6)};

\legend{HD-QP27, HD-QP37, FHD-QP27, FHD-QP37} 
\end{axis}  
\end{tikzpicture}
%\end{tabular}

%\caption{} %A caption inside a tabular section may cause errors
\begin{tikzpicture}  
\begin{axis}  
[   title={ESoC2},
    width=0.45\textwidth,height=50mm,
    ymin=0, ymax=21,
    ybar, % configures ‘bar shift’
    ymajorgrids=true,
    grid style=dashed,
    xtick=data,
    ytick={0,5,10,15,20,25},
    tick label style={font=\footnotesize},
    bar width=8pt,% height=50mm,
    xlabel={Thread configuration},%anchor=east,
    ylabel={Energy (Joule)},ylabel style={xshift=-35pt,yshift=-15pt,anchor=west},
    xlabel={Thread configuration},%anchor=east,
    legend style={at={(0.5,-0.36)},
	anchor=north,legend columns=-1},
    %legend style={at={(0.12,0.95)}, anchor=north, {nodes={scale=0.8, transform shape}}},
    %ylabel={\% load per block}, % there should be no line gap between the rows here. Otherwise, latex will show an error.  
    symbolic x coords={OpenVVC: f1/t4, f2/t2, f2/t4, VVdeC},  
    nodes near coords,  
    every node near coord/.append style={rotate=90, xshift=0cm, yshift=0cm, anchor=west,font=\footnotesize},
    nodes near coords align={vertical},  
    ] 
\addplot coordinates {(OpenVVC: f1/t4, 16.4) (f2/t2, 16.1) (f2/t4, 16.5) (VVdeC, 13.4)};%10^3
\addplot coordinates {(OpenVVC: f1/t4, 14.4) (f2/t2, 14.1) (f2/t4, 14.5) (VVdeC, 11.6)};
% \addplot coordinates {(f1/t4, 16370) (f2/t2, 16079) (f2/t4, 16475) (e8, 13393)};
% \addplot coordinates {(f1/t4, 14373) (f2/t2, 14119) (f2/t4, 14486) (e8, 11604)};

\legend{HD-QP27, HD-QP37}   
\end{axis}  
\end{tikzpicture}
\caption{Average energy (J) consumed for QP 27 and 37 sequences on ESoC1 (top) and ESoC2 (bottom) platforms.}
\label{xnene}
\end{figure}

\subsubsection{Energy consumption}
Energy consumption is another important factor for video processing operation over embedded platforms. In this study, the energy consumption was calculated as follows: 1) the power consumption is taken (in mW) after decoding each frame using the built-in power monitor of both \gls{esoc}, 2) the average power consumption of the entire sequence is multiplied by the total time in second spent for decoding the sequence, and 3) convert the energy consumption to Joule. The average energy consumption in J with different configurations of OpenVVC and \gls{vvdec} decoders is shown in Fig. \ref{xnene}. Here, OpenVVC and \gls{vvdec} have consumed comparable average energy over \gls{esoc}1 for all configurations. \gls{vvdec} has consumed on average $\times 1.17$ higher energy for \gls{hd} sequences and $\times 1.04$ higher energy for \gls{fhd} sequences with respect to the f4/t2 configuration of OpenVVC consumption on \gls{esoc}1 platform. Similar to the implementation over \gls{esoc}1, the f2/t2 configuration has used the least amount of average energy over \gls{esoc}2. Moreover, the average energy consumption of OpenVVC is slightly higher than \gls{vvdec} consumption over \gls{esoc}2.

% The energy consumption was calculated by following: 1) the power consumption is taken (in mW) after decoding each frame using build in power monitor of the both ESoCs, 2) the average power consumption of entire sequence is multiplied by the total time spend for decoding the sequence (in sec), and 3) convert the energy consumption to Joule.       
% needed in second column of first page if using \IEEEpubid
%\IEEEpubidadjcol

\subsubsection{Comparison between OpenVVC and VVdeC decoders}
Both open-source optimized video decoders OpenVVC and \gls{vvdec} have reached real-time for \gls{fhd} and \gls{hd} sequences over \gls{esoc}1 using 8 cores. In addition, both solutions present results near to real-time performance for \gls{uhd} sequences on \gls{esoc}1 platform. Moreover, OpenVVC and \gls{vvdec} decoders achieved an average of 22 \gls{fps} for \gls{qp}27 and 28.5 \gls{fps} for \gls{qp}37 using 8 cores. Table \ref{tab:xfps} and Table \ref{tab:nfps} show the average performance (in \gls{fps}) of OpenVVC and \gls{vvdec} using different number of threads on \gls{esoc}1 and \gls{esoc}2, respectively. The complexity overhead of OpenVVC with respect to  \gls{vvdec} is around 3\% for \gls{uhd}, 5\% for \gls{fhd} and 12\% for \gls{hd} sequences in both platforms.

\begin{table}[t]
\caption{Average performance (\gls{fps}) of OpenVVC and VVdeC decoders on ESoC1 platform with 1 and 8 cores.}
\label{tab:xfps}
\resizebox{\columnwidth}{!}{
\centering
\begin{tabular}{l|cc|cc|cc}
\toprule
          & \multicolumn{2}{c}{VVdeC (\gls{fps})} & \multicolumn{2}{|c}{OpenVVC (\gls{fps})} & \multicolumn{2}{|c}{VVdeC/OpenVVC} \\ \midrule
\# cores      & 1          & 8          & 1          & 8          & 1                & 8                \\ \midrule
UHD:QP27 & 3.17       & 21.72      & 2.94       & 21.95      & 108\%             & 99\%            \\
UHD:QP37 & 4.02       & 28.52      & 3.93       & 28.59      & 102\%             & 100\%            \\ \midrule
FHD:QP27 & 13.94      & 88.94      & 12.72      & 87.64      & 110\%             & 101\%             \\
FHD:QP37 & 16.66      & 115.62     & 16.92      & 114.15     & 98\%            & 101\%             \\ \midrule
HD:QP27   & 52.81      & 284.53     & 47.45      & 270.18     & 111\%            & 105\%             \\
HD:QP37   & 61.14      & 335.90     & 55.39      & 328.85     & 110\%            & 102\%             \\ \bottomrule
\end{tabular}
}
\end{table}

\begin{table}[t]
\caption{Average performance (\gls{fps}) of OpenVVC and VVdeC decoders on ESoC2 platform with 1 and 4 cores.}
\label{tab:nfps}
\resizebox{\columnwidth}{!}{
\centering
\begin{tabular}{l|cc|cc|cc}
\midrule
          & \multicolumn{2}{c}{VVdeC (\gls{fps})} & \multicolumn{2}{|c}{OpenVVC (\gls{fps})} & \multicolumn{2}{|c}{VVdeC/OpenVVC} \\ \midrule
\# cores      & 1          & 4          & 1          & 4          & 1                 & 4               \\ \midrule
UHD:QP27 & 1.20       & 4.29       & 1.15       & 4.20       & 104\%              & 102\%            \\
UHD:QP37 & 1.51       & 5.49       & 1.49       & 5.43       & 101\%              & 101\%            \\ \midrule
FHD:QP27 & 5.29       & 18.36      & 4.93       & 17.92      & 107\%              & 102\%            \\
FHD:QP37 & 6.81       & 23.85      & 6.43       & 23.31      & 106\%              & 102\%            \\ \midrule
HD:QP27   & 21.31      & 69.74      & 18.12      & 63.83      & 118\%             & 109\%            \\
HD:QP37   & 24.72      & 80.78      & 21.43      & 75.63      & 115\%             & 107\%            \\ \bottomrule
\end{tabular}
}
\end{table}

To summarize, there are three important parameters to take into consideration to select a video decoder for an embedded platform with limited hardware resources: the performance (fps), the energy consumed to decode a video and the memory used. The performance of the decoders compared in this paper (\gls{vvdec} and OpenVVC) is very similar and only a small improvement is achieved in \gls{vvdec} while the number of cores remains low. The energy consumed to decode a sequence is also very similar between both decodes. Finally, OpenVVC consumes less memory than \gls{vvdec} with a factor higher than $\times 2.11$. This significant reductions makes OpenVVC decoder as the best option to implement a VVC video decoder conformance in a multi core platform with limited resources.

%the best configuration of OpenVVC is f4/t2 on \gls{esoc}1 and f2/t2 on \gls{esoc}2 for average performance (in \gls{fps}) with the cost of consuming more memory than other configurations. Moreover, the performance results for f4/t2 and f2/t2 of OpenVVC and \gls{vvdec} are similar for all configurations, resolutions and \gls{qp} values over \gls{esoc}1 and \gls{esoc}2, respectively. In addition, \gls{vvdec} have consumed average $\times 1.11$ energy over \gls{esoc}1 compared to  OpenVVC configuration of f4/t2 and $\times 0.83$ energy over \gls{esoc}2 compared to  OpenVVC configuration of f2/t2. However, \gls{vvdec} has used $\times2.74$ memory compared to OpenVVC configuration of f4/t2 on \gls{esoc}1 and $\times2.96$ memory compared to  OpenVVC configuration of f2/t2 on \gls{esoc}2. Therefore, it can be concluded that OpenVVC is the lightest decoder among the two \gls{vvc} decoders that achieves real-time decoding for \gls{fhd}/\gls{hd} sequences over two resource-constrained embedded platforms.    

%\newpage 
\section{Conclusion} \label{conclusion}
This paper presents two open source \gls{vvc} decoders: OpenVVC and \gls{vvdec}, optimized for low-cost resource-constrained embedded platforms. Here, OpenVVC and \gls{vvdec} have been optimized at the level of data processing using \gls{simd} operations. In addition, tile and frame based parallelizations have been implemented in OpenVVC. Both decoders have achieved 15 to 34 respectively \gls{fps} for \gls{uhd} sequences with \gls{qp} 27 and 37, and achieved real-time decoding for all configurations of \gls{fhd} and \gls{hd} sequences over \gls{esoc}1 using 8 cores. Furthermore, 16 to 28 \gls{fps} have been obtained for \gls{fhd} sequences for \gls{qp}s 27 and 37, and real-time  decoding has been obtained for all \gls{hd} sequences by OpenVVC and \gls{vvdec} on \gls{esoc}2 using 4 cores. 
Moreover, the experimental results for the two most important factors of the embedded platform: the average energy consumption and maximum memory usage by both decoders were presented for \gls{esoc}1 and \gls{esoc}2. \gls{vvdec} has consumed on average $\times 2.74$ and $\times 2.96$ memory compared to the OpenVVC f4/t2 configuration on \gls{esoc}1 and the f2/t2 configuration on \gls{esoc}2, respectively. For average energy usages, \gls{vvdec} consumed on average $\times 1.11$ energy with respect to the OpenVVC f4/t2 configuration on \gls{esoc}1 and $\times 0.83$ energy with respect to the OpenVVC f2/t2 configuration on \gls{esoc}2.

% biography section
% 
% If you have an EPS/PDF photo (graphicx package needed) extra braces are
% needed around the contents of the optional argument to biography to prevent
% the LaTeX parser from getting confused when it sees the complicated
% \includegraphics command within an optional argument. (You could create
% your own custom macro containing the \includegraphics command to make things
% simpler here.)
%\begin{IEEEbiography}[{\includegraphics[width=1in,height=1.25in,clip,keepaspectratio]{mshell}}]{Michael Shell}
% or if you just want to reserve a space for a photo:

%\begin{IEEEbiography}{Michael Shell}
%Biography text here.
%\end{IEEEbiography}

% if you will not have a photo at all:
%\begin{IEEEbiographynophoto}{John Doe}
%Biography text here.
%\end{IEEEbiographynophoto}

% insert where needed to balance the two columns on the last page with
% biographies
%\newpage

%\begin{IEEEbiographynophoto}{Jane Doe}
%Biography text here.
%\end{IEEEbiographynophoto}

% You can push biographies down or up by placing
% a \vfill before or after them. The appropriate
% use of \vfill depends on what kind of text is
% on the last page and whether or not the columns
% are being equalized.

%\vfill

% Can be used to pull up biographies so that the bottom of the last one
% is flush with the other column.
%\enlargethispage{-5in}

% that's all folks

\begin{thebibliography}{1}

\bibitem{vvcwiki} 
"Fraunhofer HHI is proud to present the new state-of-the-art in global video coding: H.266/VVC brings video transmission to new speed," [Online]. Available: \url{https://newsletter.fraunhofer.de/-viewonline2/17386/465/11/14SHcBTt/V44RELLZBp/1}.

\bibitem{fvvc}
A. Wieckowski et al., "Towards A Live Software Decoder Implementation For The Upcoming Versatile Video Coding (VVC) Codec," 2020 IEEE International Conference on Image Processing (ICIP), 2020, pp. 3124-3128. %doi: 10.1109/ICIP40778.2020.9191199.

\bibitem{hevc2013}  
"High Efficiency Video Coding," Recommendation ITU–T H.265; 2013.

\bibitem{VVCComplexity} 
C. Feldmann, "Versatile Video Coding hits major milestone," [Online]. Available: \url{https://bitmovin.com/compression-standards-vvc-2020}.

\bibitem{vtm_ref}
"VVC test model," [Online]. Available:
\url{https://mpeg.chiariglione.org/standards/mpeg-i/versatile-video-coding/}

\bibitem{SIMD_other3}
C. C. Chi, M. Alvarez-Mesa, B. Bross, B. Juurlink and T. Schierl, "SIMD Acceleration for HEVC Decoding," in IEEE Trans. on Circuits and Systems for Video Technology, vol. 25, no. 5, pp. 841-855, May 2015. %doi: 10.1109/TCSVT.2014.2364413.

\bibitem{frame}
T. Amestoy, W. Hamidouche, C. Bergeron and D. Menard, "Quality-Driven Dynamic \gls{vvc} Frame Partitioning for Efficient Parallel Processing," 2020 IEEE International Conference on Image Processing (ICIP), 2020, pp. 3129-3133.% doi: 10.1109/ICIP40778.2020.9190928.

\bibitem{slice}
S. Gudumasu, S. Bandyopadhyay, and Y. He, "Software-based versatile video coding decoder parallelization," in Proceedings of the 11th ACM Multimedia Systems Conference (MMSys '20), Association for Computing Machinery, New York, NY, USA, 202–212, 2020. %doi: 10.1145/3339825.3391871.
\bibitem{tile}
Koziri, Maria and Papadopoulos, Panos K. and Tziritas, Nikos and Dadaliaris, Antonios N. and Loukopoulos, Thanasis and Khan, Samee U. and Xu, Cheng-Zhong, "Adaptive Tile Parallelization for Fast Video Encoding in HEVC," 2016 IEEE International Conference on Internet of Things (iThings) and IEEE Green Computing and Communications (GreenCom) and IEEE Cyber, Physical and Social Computing (CPSCom) and IEEE Smart Data (SmartData), 2016, pp. 738-743. %doi: 10.1109/iThings-GreenCom-CPSCom-SmartData.2016.156.

\bibitem{VVCstatus}
Gary Sullivan, "Deployment status of the VVC standard," ISO/IEC JTC1/SC29/WG11 JVET document Y0021 (JVET-Y0021), Teleconference, January, 2022.


\bibitem{openvvc} 
"OpenVVC software repository," [Online]. Available: \url{https://github.com/OpenVVC/OpenVVC}.

\bibitem{openvvc1} 
Thomas Amestoy, Pierre-loup Cabarat, Guillaume Gautier, Wassim Hamidouche and Daniel Menard, "OpenVVC: a Lightweight Software Decoder for the Versatile Video Coding Standard," arXiv preprint arXiv:2205.12217, 2022.

\bibitem{vvdec} 
"Fraunhofer HHI VVdeC software repository," [Online]. Available: \url{https://github.com/fraunhoferhhi/vvdec}.

\bibitem{cabac}
D. Karwowski, "Precise Probability Estimation of Symbols in VVC CABAC Entropy Encoder," in IEEE Access, vol. 9, pp. 65361-65368, 2021. %doi: 10.1109/ACCESS.2021.3075875.

\bibitem{mts7} 
M. J. Garrido, F. Pescador, M. Chavarrías, P. J. Lobo, C. Sanz and P. Paz, "An FPGA-Based Architecture for the Versatile Video Coding Multiple Transform Selection Core," in IEEE Access, vol. 8, pp. 81887-81903, 2020. %doi: 10.1109/ACCESS.2020.2991299.

\bibitem{ilfnst} 
Z. Hong, J. Lin, D. Jiang and J. Yin, "Improve the Efficiency of Low Frequency Non-Separable Secondary Transform Based on Implicit Multiple Transform Selection," 2019 International Conference on Artificial Intelligence and Advanced Manufacturing (AIAM), 2019, pp. 148-151. %doi: 10.1109/AIAM48774.2019.00037.

\bibitem{cclm}
R. Ghaznavi-Youvalari and J. Lainema, "Joint Cross-Component Linear Model For Chroma Intra Prediction," 2020 IEEE 22nd International Workshop on Multimedia Signal Processing (MMSP), 2020, pp. 1-5. %doi: 10.1109/MMSP48831.2020.9287167.

\bibitem{alfg} 
Y. Li et al., "An Optimized H.266/VVC Software Decoder On Mobile Platform," 2021 Picture Coding Symposium (PCS), 2021, pp. 1-5. %doi: 10.1109/PCS50896.2021.9477484.

\bibitem{vvcom}
F. Bossen, K. Sühring, A. Wieckowski and S. Liu, "VVC Complexity and Software Implementation Analysis," in IEEE Transactions on Circuits and Systems for Video Technology, vol. 31, no. 10, pp. 3765-3778, Oct. 2021. %doi: 10.1109/TCSVT.2021.3072204.

\bibitem{dmvr} 
D. Fedorov. "Decoder-Side Motion Vector Refinement (DMVR) in VVC," July. 2021 [Online]. Available: \url{https://vicuesoft.com/blog/titles/DMVR_in_VVC/}. 

\bibitem{bdof} 
D. fedorov. "Bi-directional Optical Flow (BDOF) Prediction Refinement in VVC," Oct. 2021 [Online]. Available: \url{https://vicuesoft.com/blog/titles/bi_directional_optical_flow_bdof_prediction_refinement_in_vvc/}. 

\bibitem{lmcs}
T. Lu et al., "Luma Mapping with Chroma Scaling in Versatile Video Coding," 2020 Data Compression Conference (DCC), 2020, pp. 193-202. %doi: 10.1109/DCC47342.2020.00027.

\bibitem{vil}
M. Karczewicz et al., "VVC In-Loop Filters," in IEEE Transactions on Circuits and Systems for Video Technology, vol. 31, no. 10, pp. 3907-3925, Oct. 2021, doi: 10.1109/TCSVT.2021.3072297.

\bibitem{comv}
A. Saha, M. Chavarrías, F. Pescador, Á.M. Groba, K. Chassaigne, P.L. Cebrián, "Complexity Analysis of a Versatile Video Coding Decoder over Embedded Systems and General Purpose Processors," Sensors 2021, 21, 3320. %https://doi.org/10.3390/s21103320.

\bibitem{armneon} 
“ARM Developer, Neon,” [Online]. Available: \url{https://developer.arm.com/architectures/instruction-sets/simd-isas/neon}.

\bibitem{til}
B. Bross et al., "Overview of the Versatile Video Coding (VVC) Standard and its Applications," in IEEE Transactions on Circuits and Systems for Video Technology, vol. 31, no. 10, pp. 3736-3764, Oct. 2021. %doi: 10.1109/TCSVT.2021.3101953.

\bibitem{lyjz}
L. Yan, Y. Duan, J. Sun and Z. Guo, "Implementation of HEVC decoder on x86 processors with SIMD optimization," 2012 Visual Communications and Image Processing, 2012, pp. 1-6. %doi: 10.1109/VCIP.2012.6410845.

\bibitem{uls1}
D. F. de Souza, A. Ilic, N. Roma and L. Sousa, "HEVC in-loop filters GPU parallelization in embedded systems," 2015 International Conference on Embedded Computer Systems: Architectures, Modeling, and Simulation (SAMOS), 2015, pp. 123-130. %doi: 10.1109/SAMOS.2015.7363667.

\bibitem{uls2}
D. F. de Souza, A. Ilic, N. Roma and L. Sousa, "GPU-assisted HEVC intra decoder," Journal of Real-Time Image Processing, 2016, vol. 12, Issue 2, pp. 531- 547. %doi: 10.1007/s11554-015-0519-1. 

\bibitem{gpuxssw}
X. Han, S. Wang, S. Ma and W. Gao, "Optimization Of Motion Compensation Based On GPU And CPU For VVC Decoding," 2020 IEEE International Conference on Image Processing (ICIP), 2020, pp. 1196-1200. %doi: 10.1109/ICIP40778.2020.9190708.

\bibitem{binZhu}
B. Zhu et al., "A Real-Time H.266/VVC Software Decoder," 2021 IEEE International Conference on Multimedia and Expo (ICME), 2021, pp. 1-6, doi: 10.1109/ICME51207.2021.9428470.

\bibitem{o266}
“Tencent O266dec decoder library,” [Online]. Available: \url{https://github.com/TencentCloud/O266player}.

\bibitem{vlc}
"VLC media player: VideoLAN, a project and a non-profit organization," [Online]. Available: \url{https://www.videolan.org/}.

\bibitem{gpac}
"GPAC: Multimedia Open Source Project," [Online]. Available: \url{https://gpac.wp.imt.fr/}.

\bibitem{ffplay}
"Ffmpeg: A complete, cross-platform solution to record, convert and stream audio and video," [Online]. Available: \url{https://ffmpeg.org/}.

\bibitem{2vvdec} 
J. Boyce, E. Alshina, F. Bossen, K. Kawamura, I. Moccagatta and W. Wan, "Conformance testing for versatile video coding (Draft 6)," Doc. JVET-U2008 of ITUT/ISO/IEC Joint Video Experts Team (JVET), 21st JVET meeting: January 2021.

\bibitem{3vvdec}
A. Wieckowski, C. Lehmann, B. Bross, D. Marpe, T. Biatek, M. Raulet, and J. Le Feuvre, "A Complete End to End Open Source Toolchain for the Versatile Video Coding (VVC) Standard," 2021 Proceedings of the 29th ACM International Conference on Multimedia, Association for Computing Machinery, New York, NY, USA, 3795–3798. %DOI:https://doi.org/10.1145/3474085.3478320

\bibitem{Amestoy} 
T. Amestoy, W. Hamidouche, C. Bergeron and D. Menard, "Quality-Driven Dynamic VVC Frame Partitioning for Efficient Parallel Processing," 2020 IEEE International Conference on Image Processing (ICIP), 2020, pp. 3129-3133. %doi: 10.1109/ICIP40778.2020.9190928.
\bibitem{simde} 
E. Nemerson, "Transitioning SSE/AVX code to NEON with SIMDe," [Online]. Available: \url{https://simd-everywhere.github.io/blog/2020/06/22/transitioning-to-arm-with-simde.html}.

\bibitem{xavier} 
"NVIDIA Jetson AGX Xavier Developer Kit, User Guide," DA\_09403\_003, December 17, 2019, [Online]. Available: \url{https://developer.nvidia.com/jetson-agx-xavier-developer-kit-user-guide}.

\bibitem{nano} 
"NVIDIA Jetson Nano Developer Kit, User Guide," DA\_09402\_004, January 15, 2020, [Online]. Available: \url{https://developer.nvidia.com/embedded/dlc/Jetson_Nano_Developer_Kit_User_Guide}.

\bibitem{testset} 
F. Bossen, J. Boyce, X. Li, V. Seregin, and K. Sühring, "JVET Common Test Conditions and Software Reference Configurations for SDR Video," Document JVET-N1010, JVET of ITU-T, Geneva, Mar 2019.


% references section

% can use a bibliography generated by BibTeX as a .bbl file
% BibTeX documentation can be easily obtained at:
% http://mirror.ctan.org/biblio/bibtex/contrib/doc/
% The IEEEtran BibTeX style support page is at:
% http://www.michaelshell.org/tex/ieeetran/bibtex/
%\bibliographystyle{IEEEtran}
% argument is your BibTeX string definitions and bibliography database(s)
%\bibliography{IEEEabrv,../bib/paper}
%
% <OR> manually copy in the resultant .bbl file
% set second argument of \begin to the number of references
% (used to reserve space for the reference number labels box)

\end{thebibliography}
\end{document}